\documentclass[12pt]{article}
\usepackage{a4wide,amssymb,cite}
\pdfoutput=1

\usepackage{a4wide,amssymb,graphicx}
\usepackage{epsfig}
\usepackage[usenames,dvipsnames]{color}
\usepackage{slashed}
\usepackage{amssymb,cite,graphicx}
\usepackage{slashed}
\usepackage{amsmath,bm,bbm}
\usepackage{amsfonts}
\usepackage[titletoc,title]{appendix}
\usepackage[small]{caption}
\usepackage[margin=1in]{geometry}
\usepackage[multiple]{footmisc}
\usepackage{mathtools}
\usepackage{slashed} 
\usepackage[nottoc]{tocbibind}
\usepackage{xcolor}
\usepackage{subcaption}

\newcommand{\be}{\begin{equation}}
\newcommand{\ee}{\end{equation}}
\newcommand{\bea}{\begin{eqnarray}}
\newcommand{\eea}{\end{eqnarray}}

\def\circa#1{\,\raise.3ex\hbox{$#1$\kern-.75em\lower1ex\hbox{$\sim$}}\,}

\begin{document}

\begin{titlepage}
%
%


%

\begin{centering}
\vspace{1cm}
{\Large {\bf Muon $g-2$ and Proton Lifetime in \vspace{0.2cm} \\ SUSY SU(5) GUTs with Split Superpartners }} \\

\vspace{1.5cm}

{\bf Seong-Sik Kim$^{\ddagger}$, Hyun Min Lee$^{\dagger}$, and Sung-Bo Sim$^{\sharp}$}
\\
\vspace{.5cm}

{\it Department of Physics, Chung-Ang University, Seoul 06974, Korea.}

\vspace{.5cm}


\end{centering}
\vspace{2cm}

\begin{abstract}
\noindent
We consider the interplay of the muon $g-2$ anomaly and the proton decay in the SUSY SU(5) GUTs with generation-independent scalar soft masses. In these scenarios, we introduce a number of $\bf 5+{\bar 5}$ messenger fields with doublet-triplet splitting in general gauge mediation to transmit SUSY breaking to the visible sector by gauge loops. As a result, squarks and sleptons receive generation-independent soft SUSY breaking masses, which are split already at the messenger scale. Taking into account the perturbative unification of gauge couplings as well as the bounds from electroweak precision and vacuum stability bounds, we showed the parameter space in general gauge mediation to explain the muon $g-2$ anomaly with smuon and sneutrino loops while evading the strong bounds on squarks and gluinos from the Large Hadron Collider. We also obtained the dominant Higgsino contributions to the proton decay mode, $p\to K^+{\bar\nu}$, with general generation-independent sparticle masses for squarks and sleptons. Even for split scalar soft masses in our model, however, we found that the bounds from the proton decay are satisfied only if the effective Yukawa couplings of the colored Higgsinos are suppressed further by a factor of order $10^{-4}-10^{-3}$. We illustrated how such a suppression factor is realized in orbifold GUTs in the extra dimension where the colored Higgsinos in the bulk are not coupled to the matter fields localized at the orbifold fixed points at the leading order.

\end{abstract}

\vspace{3cm}

\begin{flushleft} 
$^\ddagger$Email: sskim.working@gmail.com \\   
$^\dagger$Email: hminlee@cau.ac.kr \\
${}^{\sharp}$Email: ss971023@gmail.com 
\end{flushleft}

\end{titlepage}

\section{Introduction}

Supersymmetry (SUSY) \cite{susy} provides an elegant solution to the gauge hierarchy problem in the Standard Model (SM) due to the cancellation of huge radiative corrections to the Higgs mass with supersymmetric particles (sparticles) at the weak scale or TeV scale \cite{hierarchy,unification}. Moreover, three gauge couplings in the SM become unified at high energy due to the Renormalization Group (RG) running with common sparticle masses at low energy \cite{unification} whereas the lightest neutral superpartner is a stable dark matter candidate for Weakly Interacting Massive Particles (WIMP) in the presence of $R$-parity. However, since the Large Hadron Collider (LHC) turned on, there have been no evidence for sparticles and the limits on the colored sparticle masses such as squarks and gluinos have reached about $2\,{\rm TeV}$ or beyond \cite{gluino,squarks,CMSSUSY}. Weak-scale electroweak superpartners such as neutralinos/charginos and sleptons have been also constrained significantly \cite{CMSSUSY,sleptonsearch,LLP}, but not ruled out completely.

Recently the new measurement of the muon $g-2$ at Fermilab \cite{fermilab-run1,fermilab-run2}  has confirmed the old measurement at Brookhaven \cite{brookhaven} and strengthened the tension between the SM prediction and the experiments, so models for new physics explaining the muon $g-2$ anomaly have drawn a new attention, although we need to understand better the hadronic contributions to the muon $g-2$ within the SM.
In particular, in the context of SUSY, it is timely to look for the signatures of electroweak superpartners at the LHC in connection to the muon $g-2$ anomaly and build up a consistent picture for split masses for squarks/gluinos and the rest superparticles in concrete SUSY mediation scenarios.

As the idea of gauge coupling unification is reinforced in SUSY models,  it is natural to embed the Minimal Supersymmetric Standard Model (MSSM) into Grand Unified Theories (GUTs). In this regard, the discovery of proton decay will play a decisive role of unraveling the hints of unified theories by connecting between the unification scale and the low-energy rare processes. In the minimal SU(5) GUT, the dimension-6 operators originated from the $X$ and $Y$ GUT gauge bosons are responsible for proton decays such as $p\to \pi^0 e^+$, which can be consistent with the current bounds on the proton lifetime, $\tau(p\to \pi^0 e^+)>2.4\times 10^{34}\,{\rm years}$ \cite{SK0}. However, the colored Higgsinos within $5$ and ${\bar 5}$ Higgs multiplets lead to baryon number violating dimension-5 operators with SM quarks/leptons and scalar superpartners, so MSSM gaugino \cite{gaugino} or Higgsino loops \cite{higgsino} would induce dangerous proton decays through $p\to K^+{\bar\nu}$ if superparticle masses are heavy enough \cite{mSU5}. However, it is remarkable that scalar superparner and Higgsino masses of order $10^2\,{\rm TeV}$ \cite{SU5,update} increase proton lifetime to the level of the current experimental bounds, $\tau(p\to K^+{\bar\nu})>6.6\times 10^{33}\,{\rm years}$ \cite{SK}. Thus, it is important to check proton decay processes for the consistency of SUSY GUTs with general split sparticle masses. From the effective theory point of view below the Planck scale, the problem with dimension-5 operators is severe, because the dimensionless coefficients of such dimension-5 operators must be suppressed sufficiently, unless there is a symmetry protection mechanism for them such as discrete $R$ symmetries \cite{Rsym}.

In this article, we revisit the effects of sleptons and electroweak superpartners on the muon $g-2$ in MSSM and the proton lifetime via $p\to K^+{\bar\nu}$ in the SUSY SU(5) GUTs with general messenger fields. To this purpose, we introduce the general formula for the muon $g-2$ with new spin-0 or spin-$\frac{1}{2}$ charged or neutral particles in loops.  We also consider general gauge mediation \cite{gaugemed,gauge-review} for SUSY breaking mass parameters for superpartners and show that squarks and gluinos can be much heavier than sleptons and electroweak superpartners, and scalar soft masses are generation-independent. Anomaly-gravity mixed mediation can be an option to realize the split spectrum for sparticles, with a fine-tuning of soft mass parameters of different origin. 

Assuming doublet-triplet splitting masses in the messenger chiral multiplets in the representation of $\bf 5+{\bar 5}$ under $SU(5)$ \cite{EOGM,EOGM2,EOGM-amm}, we parametrize the scalar soft masses for squarks and sleptons in terms of the effective numbers of colored and non-colored messengers, $N_3$ and $N_2$, respectively, and the messenger scale, $\Lambda_G$. As a result, taking into account electroweak precision and vacuum stability bounds, we identify the parameter space for explaining the muon $g-2$ anomaly and show whether the muon $g-2$ constraint favors the lighter slepton to be heavier or lighter than the lightest neutralino in MSSM. In gauge mediation scenarios, the gravitino, the superpartner of graviton, can be the lightest superparticle (LSP), which is lighter than the MSSM LSP, so the experimental bounds from the standard SUSY searches with missing transverse momentum are not directly applicable. We also check the unification condition with split messenger fields and split MSSM fields in general gauge mediation. 

Taking the benchmark models with split sparticle masses for explaining the muon $g-2$ anomaly and realizing squark and gluino masses satisfying the current LHC bounds, we obtain the proton lifetime from $p\to K^+{\bar\nu}$  as a function of the squark mass parameter. Having in mind the embedding of SUSY GUTs into orbifold GUTs in the extra dimension, we introduce the suppression factor $\kappa$ for the Yukawa couplings for the colored Higgsinos.
In this case, we show how the parameter space for the messenger scale $\Lambda_G $ versus the suppression factor $\kappa$ is constrained by the current bounds on the proton decay.

The paper is organized as follows. 
We first begin with the setup for the slepton interactions with gauginos and Higgsinos in MSSM and the mass matrices for neutralinos, charginos and sleptons. Then, we discuss the possibility of generation-independent but non-universal sparticle masses in general gauge mediation and mention another example with anomaly-gravity mixed mediation. We also check the consistency conditions for gauge coupling unification, electroweak precision and vacuum stability bounds. Next we provide the general formulas for the muon $g-2$ and apply them to the supersymmetric case with sleptons and look for the parameter space for explaining the muon $g-2$ anomaly in general gauge mediation. We continue to discuss the general effective Lagrangian relevant for the proton decay mode, $p\to K^+{\bar\nu}$, and obtain the general formula for the proton lifetime for generation-independent but non-universal soft mass parameters.  Then, conclusions are drawn. There is one appendix summarizing the neutralino and chargino mixings in the perturbative approximation and the effective interactions for the sleptons in the basis of mass eigenstates for sleptons and electroweak superpartners.

\section{The setup}

We introduce the slepton interactions and the mass matrices for neutralinos,  charginos and sleptons in MSSM.

\subsection{Slepton interactions}

In the flavor basis for leptons and scalar superpartners, we introduce $l_{iL}=(\nu_{iL},e_{iL})^T$ and ${\tilde l}_{iL}=({\tilde\nu}_{iL},{\tilde e}_{iL})^T$ as $SU(2)_L$ doublets and $e^c_{iR}$ and ${\tilde e}^*_{iR}$ as $SU(2)_L$ singlets. Then, we consider the interactions of the sleptons to electroweak gauginos (${\widetilde B}, {\widetilde W}^3, {\widetilde W}^\pm$) and Higgsinos (${\widetilde H}_d=({\widetilde H}^0_d,{\widetilde H}^-_d)$), as follows,
\bea
{\cal L}_{\rm sleptons} &=&-f^{ij}_l {\widetilde H}^0_d {\tilde e}_{iL} e^c_{jR}+f^{ij}_l {\widetilde H}^-_d {\tilde \nu}_{iL} e^c_{jR} -f^{ij}_l {\widetilde H}^0_d e_{iL} {\tilde e}^*_{jR} +f^{ij}_l {\widetilde H}^-_d \nu_{iL} {\tilde e}^*_{jR} +{\rm h.c.} \nonumber \\
&&-\sqrt{2} g' \Big({\tilde l}^\dagger_{iL} Y_{l_L}  {\widetilde B} l_{iL} + {\tilde e}_{iR} Y_{e^c_R} {\widetilde B}  e^c_{iR}+{\rm h.c.} \Big)-\sqrt{2}g \Big({\tilde l}^\dagger_{iL}  T^3 {\widetilde W}^3 l_{iL} +{\rm h.c.}\Big) \nonumber \\
&&-g \Big({\tilde e}^*_{iL} {\widetilde W}^- \nu_{iL} +{\tilde\nu}^*_{iL} {\widetilde W}^+ e_{iL} +{\rm h.c.} \Big) \label{sleptonint}
\eea
where $f^{ij}_l$ are the Yukawa couplings for the charged leptons, $Y_{l_L}, Y_{e^c_R}$ are the hypercharges of $l_L$ and $e^c_R$, given by $Y_{l_L}=-\frac{1}{2}$ and $Y_{e^c_R}=+1$, and $T^3=\frac{1}{2}\sigma^3$.
When the slepton masses are flavor universal, we can write the Higgsino and gaugino interactions in the diagonal form in the basis of mass eigenstates of leptons after the sleptons are simultaneously rotated. But, there is still a mixing between the superpartners of left-handed and right-handed charged leptons in each generation, such as ${\tilde e}_{iL}$ and ${\tilde e}_{iR}$, as will be discussed shortly.

\subsection{Neutralinos and charginos}

The mass terms for neutralinos, $\{{\widetilde B}, {\widetilde W}^3,{\widetilde H}^0_{d}, {\widetilde H}^0_{u}\}$, charginos, $\{{\widetilde W}^-, {\widetilde H}^-_d\}$,  and their complex conjugates, are given, respectively, by
\bea
{\cal L}_N=-\frac{1}{2} ({\widetilde B}, {\widetilde W}^3,{\widetilde H}^0_{d}, {\widetilde H}^0_{u}){\cal M}_N \left(\begin{array}{c} {\widetilde B} \\{\widetilde W}^3 \\  {\widetilde H}^0_d \\ {\widetilde H}^0_u \end{array} \right)+{\rm h.c.}
\eea
with
\bea
{\cal M}_N =\left(\begin{array}{cccc} M_1 &  0 & -m_Z\sin\theta_W \cos\beta & m_Z\sin\theta_W\sin\beta \\ 0  & M_2  & m_Z\cos\theta_W\cos\beta  & -m_Z\cos\theta_W\sin\beta \\  -m_Z\sin\theta_W \cos\beta & m_Z\cos\theta_W\cos\beta & 0  & -\mu_H  \\ m_Z\sin\theta_W \sin\beta & -m_Z\cos\theta_W\sin\beta  & -\mu_H & 0 \end{array} \right),
\eea
and
\bea
{\cal L}_C=-(\overline{{\widetilde W}^-_R}, \overline{{\widetilde H}^-_{uR}} ) {\cal M}_C \left(\begin{array}{c}{\widetilde W}^-_L  \\ {\widetilde H}^-_{dL} \end{array}\right)
\eea
with
\bea
{\cal M}_C=\left(\begin{array}{cc}  M_2 & \sqrt{2} m_W \cos\beta \\ \sqrt{2}m_W \sin\beta  & \mu_H  \end{array}\right).
\eea
Here, $M_{1,2}$ are the soft SUSY breaking masses for Bino and Wino gauginos, and $\mu_H$ is the supersymmetric mass parameter for the Higgsinos. We note that $\sin\beta=v_u/v$ and $\cos\beta=v_d/v$ with $v=\sqrt{v^2_u+v^2_d}$,  for $\langle H^0_u\rangle=\frac{1}{\sqrt{2}}v_u$ and  $\langle H^0_d\rangle=\frac{1}{\sqrt{2}}v_d$.

The mass matrix for neutralinos can be diagonalized by
\bea
N^*{\cal M}_N N^\dagger = {\cal M}^{\rm diag}_N= {\rm diag}(m_{{\widetilde \chi}^0_1},m_{{\widetilde \chi}^0_2},m_{{\widetilde \chi}^0_3},m_{{\widetilde \chi}^0_4})
\eea
where the rotation matrix $N$ defines the mass eigenstates for neutralinos in four-component spinor notations as
\bea
{\widetilde \chi}^0_{iL} &=& N_{i1} {\widetilde B}_L+N_{i2} {\widetilde W}^3_L+ N_{i3} {\widetilde H}^0_{d,L}+N_{i4} {\widetilde H}^0_{u,L}, \\
{\widetilde \chi}^0_{iR} &=& N^*_{i1} {\widetilde B}_R+N^*_{i2} {\widetilde W}^3_R+ N^*_{i3} {\widetilde H}^0_{d,R}+N^*_{i4} {\widetilde H}^0_{u,R},\label{neutralinomix}
\eea
or 
\bea
{\widetilde B}_L&=&N^*_{i1} {\widetilde \chi}^0_{iL},  \quad {\widetilde W}^3_L=N^*_{i2} {\widetilde \chi}^0_{iL}, \quad {\widetilde H}^0_{d,L}=N^*_{i3} {\widetilde \chi}^0_{iL}, \quad {\widetilde H}^0_{u,L}=N^*_{i4} {\widetilde \chi}^0_{iL}, \\
{\widetilde B}_R&=&N_{i1} {\widetilde \chi}^0_{iR},  \quad {\widetilde W}^3_R=N_{i2} {\widetilde \chi}^0_{iR}, \quad {\widetilde H}^0_{d,R}=N_{i3} {\widetilde \chi}^0_{iR}, \quad {\widetilde H}^0_{u,R}=N_{i4} {\widetilde \chi}^0_{iR}.
\eea
Similarly, the mass matrix for charginos can be also diagonalized by
\bea
U_R {\cal M}_C U^\dagger_L ={\cal M}^{\rm diag}_C= {\rm diag}(m_{{\widetilde \chi}^-_1},m_{{\widetilde \chi}^-_2}), 
\eea
with
\bea
m^2_{{\widetilde \chi}^-_1,{\widetilde \chi}^-_2} =\frac{1}{2}\bigg[|M_2|^2+|\mu_H|^2+2m^2_W \mp \sqrt{(|M_2|^2+|\mu_H|^2+2m^2_W)^2-4|\mu_H M_2-m^2_W\sin 2\beta|^2} \bigg],
\eea
and the mass eigenstates for charginos are
\bea
 \left(\begin{array}{c}{\widetilde \chi}^-_1  \\ {\widetilde \chi}^-_2 \end{array}\right)_{L,R} =U_{L,R}  \left(\begin{array}{c}{\widetilde W}^-_L  \\ {\widetilde H}^-_{dL} \end{array}\right)_{L,R}. \label{charginomix}
\eea

In the limit of small electroweak symmetry breaking effects, namely, $m_Z\ll |\mu_H\pm M_1|, |\mu\pm M_2|$, we can approximate the neutralino masses as
\bea
m_{{\widetilde \chi}^0_1} &\simeq & M_1 -\frac{m^2_Z s^2_W (M_1+\mu_H\sin 2\beta)}{\mu^2_H-M^2_1}, \\
m_{{\widetilde \chi}^0_2} &\simeq & M_2-\frac{m^2_W (M_2+\mu_H\sin 2\beta)}{\mu^2_H-M^2_2}, \\
m_{{\widetilde \chi}^0_3} &\simeq & \mu_H+ \frac{m^2_Z(1+\sin2\beta)(\mu_H-M_1 c^2_W-M_2s^2_W)}{2(\mu_H-M_1)(\mu_H-M_2)}, \\
m_{{\widetilde \chi}^0_4} &\simeq & -\mu_H  + \frac{m^2_Z(1-\sin2\beta)(\mu_H+M_1 c^2_W+M_2s^2_W)}{2(\mu_H+M_1)(\mu_H+M_2)}.
\eea
Here, for $\mu_H>0$, the leading mass for ${\widetilde \chi}^0_4$ is negative, so we need to rescale ${\widetilde \chi}^0_4$ to $i{\widetilde \chi}^0_4$ to get a positive mass for ${\widetilde \chi}^0_4$. On the other hand, for $\mu_H<0$, instead we need to rescale ${\widetilde \chi}^0_3$ to $i{\widetilde \chi}^0_3$ to get a positive mass for ${\widetilde \chi}^0_3$.
Then, in the leading order approximation with a heavy Higgsino, the neutralino mass eigenstates become ${\widetilde\chi}^0_1\simeq {\widetilde B}$ (bino-like),  ${\widetilde\chi}^0_2\simeq {\widetilde W}^3$ (wino-like), ${\widetilde\chi}^0_3, {\widetilde\chi}^0_4 \simeq ({\widetilde H}^0_u\mp  {\widetilde H}^0_d)/\sqrt{2}$ (higgsino-like). 

Similarly, $m_W\ll |\mu_H\pm M_2|$, the chargino masses are approximated to
\bea
m_{{\widetilde \chi}^-_1} &\simeq&M_2 -\frac{m^2_W (M_2+\mu_H\sin 2\beta)}{\mu^2_H-M^2_2},  \\
m_{{\widetilde \chi}^-_2} &\simeq&\mu_H+ \frac{m^2_W (\mu_H+M_2\sin 2\beta)}{\mu^2_H-M^2_2},
\eea
and chargino mass eigenstates become ${\widetilde \chi}^-_1 \simeq {\widetilde W}^-$ (wino-like), ${\widetilde \chi}^-_2 \simeq {\widetilde H}^-_d$ (higgsino-like).  For $\mu_H<0$, we need to rescale the Higgsino-like chargino by ${\widetilde \chi}^-_2\to i{\widetilde \chi}^-_2$ to get a positive mass for ${\widetilde \chi}^-_2$. We refer to Appendix A for the neutralino and chargino mixing matrices with corrections coming from the electroweak symmetry breaking.

\subsection{Charged sleptons}

The mass terms for the charged sleptons are given by
\bea
{\cal L}_{\tilde e} = -({\tilde e}^*_L, {\tilde e}^*_R) {\cal M}^2_{\tilde e} \left(\begin{array}{c} {\tilde e}_L \\ {\tilde e}_R \end{array}\right) 
\eea
with
\bea
{\cal M}^2_{\tilde e}=\left(\begin{array}{cc} m^2_{LL} & m^2_{LR} \\  (m^2_{LR})^* & m^2_{RR} \end{array}\right) \label{sleptonmass}
\eea
where 
\bea
m^2_{LL}&=&m^2_{{\tilde e}_L}+m^2_Z\cos2\beta \Big(s^2_W-\frac{1}{2}\Big)+m^2_e,  \\
m^2_{RR}&=&m^2_{{\tilde e}_R}-m^2_Z\cos2\beta \,s^2_W+m^2_e, \\
m^2_{LR}&=& m_l (A^*_l -\mu_H \tan\beta). 
\eea
Here, $m_e$ is the charged lepton mass, and we assumed that the trilinear soft SUSY breaking term takes the form in the alignment limit, ${\cal L}_{\rm trilinear}=-A^{ij}_lH^0_d {\tilde e}_{iL} {\tilde e}^*_{jR}+{\rm h.c.}$, with $A^{ij}_l=f^{ij}_l A_l$, and we dropped the generation indices for the charged sleptons in the mass matrix. We note that the $A$-terms are typically proportional to the Yukawa couplings in gravity or gauge mediation, so we can ignore them. However, for a large $\tan\beta$, the $\mu$ term can give rise to a large mixing for sleptons, because $m^2_{LR}\simeq -m_l \mu_H \tan\beta$.

After diagonalizing the slepton mass matrix in eq.~(\ref{sleptonmass}) with
\bea
 \left(\begin{array}{c} {\tilde e}_2\\ {\tilde e}_1 \end{array}\right) =\left(\begin{array}{cc} \cos\theta_{\tilde e} & \sin\theta_{\tilde e} \\ -\sin\theta_{\tilde e} & \cos\theta_{\tilde e} \end{array}\right) \left(\begin{array}{c} {\tilde e}_L \\ {\tilde e}_R \end{array}\right),  \label{sleptonmix}
\eea
we obtain the slepton mass eigenvalues as
\bea
m^2_{{\tilde e}_2, {\tilde e}_1} = \frac{1}{2} \bigg[m^2_{LL}+m^2_{RR} \pm (m^2_{LL}-m^2_{RR})\sqrt{1+\frac{4|m^2_{LR}|^2}{(m^2_{LL}-m^2_{RR})^2}} \bigg],
\eea
and the slepton mixing angle as
\bea
\tan2\theta_{\tilde e} =\frac{2m^2_{LR}}{m^2_{LL}-m^2_{RR}}.
\eea
Thus, for $-\frac{\pi}{2}<\theta_{\tilde e}<\frac{\pi}{2}$, we get
\bea
\sin2\theta_{\tilde e} = \frac{2m^2_{LR}}{m^2_{{\tilde e}_2}-m^2_{{\tilde e}_1}}. \label{sleptonmix}
\eea
For $m^2_{{\tilde e}_2}>m^2_{{\tilde e}_1}$ or $m^2_{LL}>m^2_{RR} $, we get $\sin2\theta_{\tilde e}<0$ for $m^2_{LR}<0$, which is consistent with the correct sign of the slepton contributions to the muon $g-2$, as will be discussed later.  On the other hand, for $m^2_{{\tilde e}_2}<m^2_{{\tilde e}_1}$ or $m^2_{LL}<m^2_{RR} $,  we get $\sin2\theta_{\tilde e}>0$ for $m^2_{LR}<0$, for which the slepton contributions to the muon $g-2$ are positive.

For the maximal mixing angle, $\theta_{\tilde e} =\pm\frac{\pi}{4}$, namely, for $m^2_{LR}=\pm\frac{1}{2} (m^2_{{\tilde e}_2}-m^2_{{\tilde e}_1})$ and $m^2_{RR}=m^2_{LL}$, we get  the slepton mass eigenvalues  as
\bea
m^2_{{\tilde e}_{2}} &= & m^2_{LL}+|m^2_{LR}|\simeq m^2_{{\tilde e}_L}+m^2_Z\cos2\beta \Big(s^2_W-\frac{1}{2}\Big)+m_l|\mu_H| \tan\beta,\\
m^2_{{\tilde e}_{1}} &= & m^2_{LL}-|m^2_{LR}| \simeq m^2_{{\tilde e}_L}+m^2_Z\cos2\beta \Big(s^2_W-\frac{1}{2}\Big)-m_l |\mu_H| \tan\beta, 
\eea
On the other hand, for a vanishing mixing angle, $|\theta_{\tilde e}|\ll 1$, we obtain $m^2_{{\tilde e}_{2}}\simeq m^2_{LL}$ and $m^2_{{\tilde e}_{1}} \simeq m^2_{RR}$.

\section{Non-universal sparticle masses}

In this section, we pursue the possibility of generation-independent but non-universal sparticle masses for squarks and sleptons at the messenger scale. 
Thus, we consider two concrete mediation mechanisms for SUSY breaking, namely, general gauge mediation and anomaly-gravity mixed mediation. 
We also discuss the RG running of gauge couplings and soft mass parameters in general gauge mediation.

\subsection{General gauge mediation}

The soft SUSY breaking masses for squarks and sleptons must be flavor diagonal in order to satisfy the bounds on Flavor Changing Neutral Currents (FCNCs), unless the supersymmetric particles are sufficiently heavy, for instance, by their masses of order $10\,{\rm TeV}$. 
For the sparticle masses that are invariant under the SM gauge groups, sparticle masses satisfy $m^2_{{\tilde u}_{iL}}=m^2_{{\tilde d}_{iL}}$ and $m^2_{{\tilde e}_{iL}}=m^2_{{\tilde \nu}_{iL}}$ above the electroweak scale, but $m^2_{{\tilde u}_{iR}}$, $m^2_{{\tilde d}_{iR}}$ and $m^2_{{\tilde e}_{iR}}$ can be different from the other sparticle masses.  

Suppose that soft SUSY breaking masses are generated by gauge mediation \cite{gaugemed} at the GUT scale in the minimal SU(5).
Then, the sparticle masses are generation-independent and they are further constrained by the $SU(5)$ unification to $m^2_{{\tilde u}_{iL}}=m^2_{{\tilde u}_{iR}}=m^2_{{\tilde d}_{iL}}=m^2_{{\tilde e}_{iR}}\equiv m^2_{\bf 10}$ and $m^2_{{\tilde d}_{iR}}=m^2_{{\tilde e}_{iL}}=m^2_{{\tilde \nu}_{iL}}\equiv m^2_{\bf {\bar 5}}$ at the unification scale \cite{gaugemed,gauge-review}. However, if there are mass splittings between the components of the messenger multiplets, for instance, due to doublet-triplet mass splitting in the case of $\bf 5+{\bar 5}$ messenger multiplets, we can generate non-universal soft masses for scalar superpartners by general gauge mediation \cite{EOGM}. 

We consider a singlet SUSY breaking chiral multiplet, $X$, and  paired messenger chiral multiplets in the $\bf 5+{\bar 5}$ representations of $SU(5)$, $\phi_i+{\widetilde \phi}_i$, with $i=1,2,\cdots, N$.
Then, after the $SU(5)$ GUT symmetry is broken,
we take the effective superpotential for the component fields of messenger chiral multiplets below the GUT scale is given by
\bea
W_{\rm eff} = (\lambda_{3,ij} X+m_{3,ij}) q_i {\widetilde q}_j +  (\lambda_{2,ij} X+m_{2,ij}) l_i {\widetilde l}_j 
\eea
where $q_i+ {\widetilde q}_i$ and $l_i + {\widetilde l}_i  $ are the component fields of the messenger multiplets. Then, taking the VEV of the SUSY breaking chiral multiplet as
\bea
\langle X \rangle =X+\theta^2 F,
\eea
we can get the mass splittings for messenger chiral multiplets, which give rise to nonzero sparticle masses in the MSSM at loops.

First, gaugino masses from general gauge mediation take the same values at the messenger scale as in the minimal gauge mediation with $SU(5)$ supersymmetric masses for messenger fields \cite{EOGM}, as follows,
\bea
M_a = \frac{\alpha_a}{4\pi}\, \Lambda_G, \quad a=1,2,3,
\eea
with $\alpha_a=g^2_a/(4\pi)$ and 
\bea
\Lambda_G =n \,\frac{F}{X},
\eea
so the running gaugino masses at low energy become
\bea
\frac{M_3}{\alpha_3}= \frac{M_1}{\alpha_1}=\frac{M_2}{\alpha_2}=\frac{\Lambda_G}{4\pi}. 
\eea
Here, the electroweak gauge couplings, $g, g'$, are related by $g_2=g$ and $g_1=\sqrt{5/3}g'$. We quote the values of the gauge couplings at the top mass scale by $g(m_t)=0.64, g'(m_t)=0.35, g_3(m_t)=1.16$. Then, we obtain the ratios of the gaugino masses at the top quark mass scale by $M_3:M_2:M_1\simeq 6:2:1$.

The soft masses for scalar superpartners in general gauge mediation are also given \cite{EOGM} by
\bea
m^2_{\tilde f} =2\sum_{a=1}^3 C_a({\tilde f}) \bigg(\frac{\alpha_a}{4\pi}\bigg)^2 \Lambda^2_a
\eea
where $C_a({\tilde f})$ are the quadratic Casimir invariants for ${\tilde f}$ under the SM gauge group, nonzero values of which are
$C_3({\tilde f})=\frac{4}{3}$ for ${\tilde f}={\tilde q}_L$, $C_2({\tilde f})=\frac{3}{4}$ for ${\tilde f}={\tilde q}_L, {\tilde l}_L$, and $C_1({\tilde f})=\frac{3}{5} Y^2_{\tilde f}$ for all scalar superpartners.
Moreover, we note that  
\bea
\Lambda^2_a =\frac{\Lambda^2_G}{N_a}, \quad a=1,2,3,
\eea
$\Lambda^2_1=\frac{2}{5} \Lambda^2_3+\frac{3}{5} \Lambda^2_2$,
with the effective number of doublet and triplet messenger fields being given by
\bea
N_a=\bigg[\frac{1}{2n^2}|X|^2\frac{\partial^2}{\partial X\partial X^*}\sum_{i=1}^{N}\bigg(\ln \frac{|{\cal M}_{a,i}|^2}{\mu^2}\bigg)^2\bigg]^{-1},\quad a=2,3. 
\eea
Here, ${\cal M}_{a,i}(a=2,3)$ are the eigenvalues of the mass matrices for colored and non-colored components of messenger fields, 
and
\bea
n=\frac{1}{R(X)}\sum_{i=1}^N (2-R(\phi_i)-R({\widetilde\phi}_i)),
\eea
is the number of messenger fields with $R(\phi_i)+R({\widetilde\phi}_i)\neq 2$, satisfying $0\leq n\leq N$,
with $R(\phi)$ being the $R$ charges of $\phi=X,\phi_i,{\widetilde\phi}_i$.
We note that $N_a$ is a continuous function of the couplings taking values between 0 and $N$ inclusive, and the asymptotic limits of $N_a$ at $X\to 0,\infty$ satisfy the following inequalities \cite{EOGM}, 
\bea
\frac{n^2}{n^2-(N-r_m-1)(2n-N+r_m)}\leq N_a (X\to 0)\leq N-r_m
\eea
and 
\bea
\frac{n^2}{r_\lambda+(r_\lambda-n)^2}\leq N_a(X\to \infty) \leq \frac{n^2}{r_\lambda+\frac{(r_\lambda-n)^2}{(N-r_\lambda)}}
\eea
where $r_\lambda\equiv {\rm rank}\,\lambda_a$ and $r_m={\rm rank}\, m_a$, satisfying $r_\lambda+r_m\geq N$ for the non-degenerate matrix, $\lambda X+m$.

More explicitly, the general gauge mediation lead to squark and slepton masses in MSSM as
\bea
m^2_{{\tilde q}_L} &=&  \frac{8}{3}\bigg(\frac{\alpha_3}{4\pi}\bigg)^2\Lambda^2_3+ \frac{3}{2}\bigg(\frac{\alpha_2}{4\pi}\bigg)^2\Lambda^2_2+\frac{1}{150}  \bigg(\frac{\alpha_1}{4\pi}\bigg)^2(2\Lambda^2_3+3\Lambda^2_2),  \\
m^2_{{\tilde u}_R}&=& \frac{8}{3}\bigg(\frac{\alpha_3}{4\pi}\bigg)^2\Lambda^2_3+\frac{8}{75}  \bigg(\frac{\alpha_1}{4\pi}\bigg)^2(2\Lambda^2_3+3\Lambda^2_2),  \\
m^2_{{\tilde d}_R}&=& \frac{8}{3}\bigg(\frac{\alpha_3}{4\pi}\bigg)^2\Lambda^2_3+\frac{2}{75}  \bigg(\frac{\alpha_1}{4\pi}\bigg)^2(2\Lambda^2_3+3\Lambda^2_2),  \\
m^2_{{\tilde l}_L} &=& \frac{3}{2}\bigg(\frac{\alpha_2}{4\pi}\bigg)^2\Lambda^2_2+\frac{3}{50}  \bigg(\frac{\alpha_1}{4\pi}\bigg)^2(2\Lambda^2_3+3\Lambda^2_2), \\
m^2_{{\tilde e}_R} &=&\frac{6}{25}  \bigg(\frac{\alpha_1}{4\pi}\bigg)^2(2\Lambda^2_3+3\Lambda^2_2).
\eea
We also remark that the scalar soft masses receive corrections, $\Delta_\phi=(T_{3\phi}-Q_\phi \sin^2\theta_W)M_Z^2\cos2\beta$, due to electroweak symmetry breaking. 

As a consequence, in general gauge mediation with $N_3\ll N_2$ or $\Lambda_3\gg \Lambda_2$, we can make the soft mass parameters for squarks much larger than those for sleptons.
Then, we can explain the deviation of the muon $g-2$ from the SM value with light sleptons, electroweak gauginos and Higgsinos \cite{EOGM-amm}, while being consistent with the LHC bounds on the squark masses. Flavor universality in general gauge mediation restricts the soft masses for sleptons to be almost generation-independent, namely, $m_{\tilde e}\sim m_{\tilde \mu}\sim m_{\tilde\tau}$, up to the RG running effects and the slepton mixings for the generation-dependent Yukawa couplings for leptons. Then, as will be discussed later, we can correlate between the muon $g-2$ with smuon in loops and the proton lifetime from $p\to K{\bar \nu}$ with stau in loops.

\begin{figure}[!t]
\centering
\includegraphics[width=0.48\textwidth,clip]{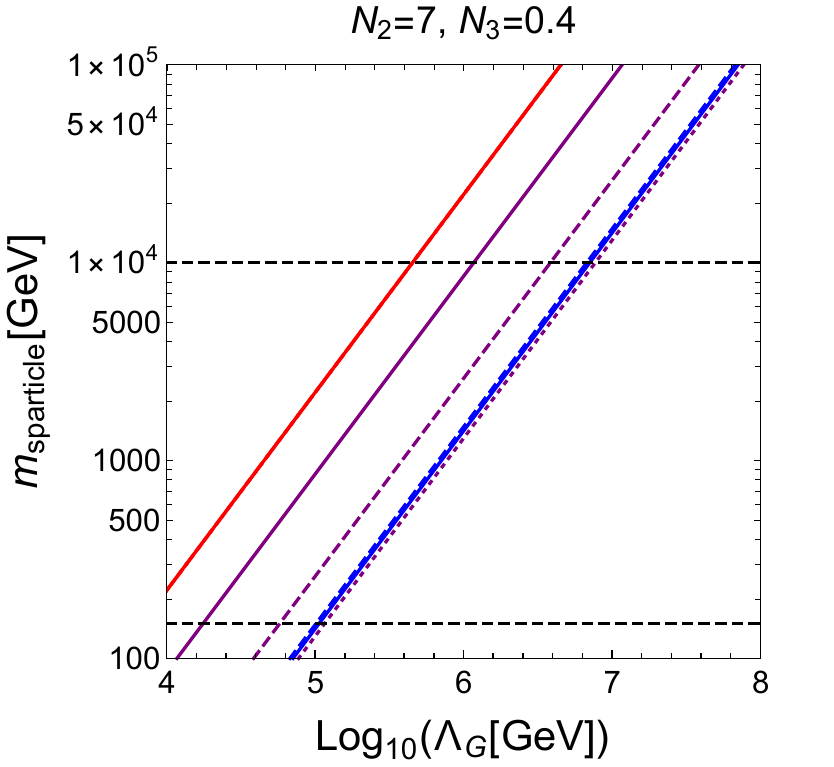}\,\,\,\,\,\,
\includegraphics[width=0.48\textwidth,clip]{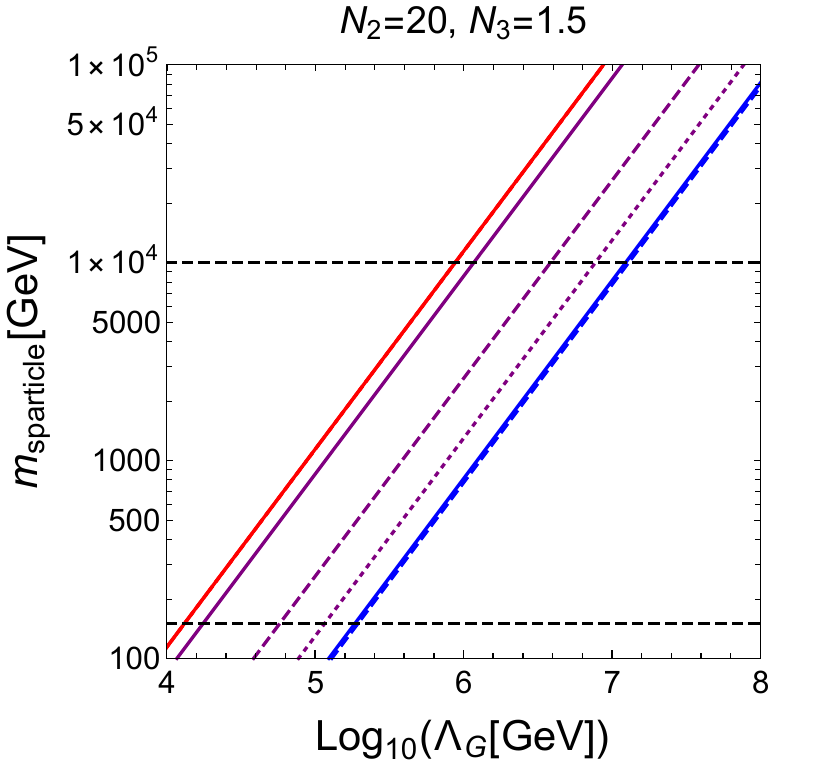} 
\caption{Soft mass parameters as a function of $\Lambda_G$ in general gauge mediation.   We showed $m_{{\tilde q}_L}$, $m_{{\tilde u}_R}$, $m_{{\tilde d}_R}$, which are almost degenerate, in red solid line, and $m_{{\tilde l}_L}$, $m_{{\tilde e}_R} $, $M_3, M_2, M_1$, in blue solid, blue dashed, purple solid, purple dashed, purple dotted lines, respectively. We chose $N_2=10, N_3=2$ on left and $N_2=20, N_3=1.5$ on right. 
}
\label{susymass}
\end{figure}

In Fig.~\ref{susymass}, we depict the soft mass parameters for squarks and sleptons in general gauge mediation as a function of $\Lambda_G$. We fixed $N_2=10, N_3=2$ on left and $N_2=20, N_3=1.5$ on right. $m_{{\tilde q}_L}$, $m_{{\tilde u}_R}$, $m_{{\tilde d}_R}$, in red solid lines, and $m_{{\tilde l}_L}$, $m_{{\tilde e}_R} $, $M_3, M_2, M_1$ are drawn in blue solid, blue dashed, purple solid, purple dashed, purple dotted lines, respectively. Here, we note that the squark masses are shown to be almost degenerate, so they are not distinguishable in the plots. We find that it is possible to introduce split masses for squarks/gluinos and electroweak superpartners such as bino, wino and sleptons in general gauge mediation. These two benchmark points will be used for a later discussion on the muon $g-2$ and the $W$ boson mass.

We comment on the $A$-terms in gauge mediation. The $A$-terms for squarks and sleptons vanish at one loop in gauge mediation, but they can be generated by the RG evolution proportional to gaugino masses \cite{gauge-review}. Thus, the $A$-terms for sleptons are suppressed by loops as compared to slepton masses, so we can ignore the contribution of the $A$-terms to the mixing between the smuons in this case.  Moreover, the $\mu$ and $B\mu$ terms call for the solution from singlet extensions of the MSSM and other SUSY mediation scenarios such as anomaly mediation \cite{EOGM2}.

\subsection{Anomaly-gravity mixed mediation}

In anomaly mediation \cite{anomalymed}, the predictive patterns for sparticle masses are given in terms of the SUSY breaking scale of the conformal compensator, $F_\Phi$, and the SM gauge interactions at low energy, but there is a problem with tachyonic slepton masses \cite{anomalymed}. However, adding the universal masses for squarks and sleptons at the GUT scale in gravity mediation, we can cure the tachyonic mass problem for sleptons  \cite{anomalymed}. In this case, when the slepton masses in anomaly mediation are almost cancelled by those from gravity mediation, we can make sleptons much lighter than squarks.

First, the gaugino masses in anomaly mediation are given by
\bea
M_a =\frac{\alpha_a}{4\pi}\, b_a\,F_\Phi, \quad a=1,2,3,
\eea
where $b_a=(-\frac{33}{5},-1,3)$ are the one-loop beta functions for the SM gauge couplings.
On the other hand, the general formula for scalar soft masses in anomaly mediation is given by
\bea
m^2_i = -\frac{1}{4} \bigg(\frac{\partial\gamma_i}{\partial g}\,\beta_g + \frac{\partial \gamma_i}{\partial y}\bigg) |F_\Phi|^2
\eea
where $\gamma_i, \beta_g$ are the anomalous dimension for scalar sparticle $i$, the beta function for the gauge coupling $g$, and $y$ is the Yukawa coupling.  Thus, adding the universal scalar soft masses, $m^2_0$, we obtain the total scalar soft masses \cite{EOGM2} as
\bea
m^2_{{\tilde q}_{3L}} &=& m^2_0+\bigg[8\alpha^2_3-\frac{3}{2} \alpha^2_2-\frac{11}{50}\alpha^2_1+\alpha^2_t \bigg(6\alpha^2_t -\frac{13}{15} \alpha_1 -3 \alpha_3 -\frac{16}{13}\alpha_3\bigg)\bigg] |F_\Phi|^2 ,  \label{anom-q} \\
m^2_{{\tilde u}_{3R}}&=&m^2_0+\bigg[8\alpha^2_3-\frac{88}{25} \alpha^2_1 +2\alpha^2_t \bigg(6\alpha^2_t -\frac{13}{15} \alpha_1 -3 \alpha_3 -\frac{16}{13}\alpha_3\bigg) \bigg] |F_\Phi|^2,  \label{anom-u}\\
m^2_{{\tilde d}_R}&=&m^2_0+\bigg(8\alpha^2_3-\frac{22}{25} \alpha^2_1 \bigg)  |F_\Phi|^2,  \\
m^2_{{\tilde l}_L} &=&m^2_0-\bigg(\frac{99}{50}\alpha^2_1+\frac{3}{2}\alpha^2_2\bigg)|F_\Phi|^2, \\
m^2_{{\tilde e}_R} &=&m^2_0 -\frac{198}{25} \alpha^2_1 |F_\Phi|^2,
\eea
with $\alpha_t=y^2_t/(4\pi)$.
The masses for the first and second generations of up-type squarks are the same as in eqs.~(\ref{anom-q}) and (\ref{anom-u}), but with $\alpha_t=0$.
As a result, choosing $m^2_0\simeq c\, \alpha^2_2 |F_\Phi|^2$ with $c$ being of order one and positive, we can make the mass squares for sleptons to be positive and much lighter than squarks.  However, we need a fine-tuning to get the slepton masses much smaller than the squark masses in this case, so we don't pursue the detailed discussion on this possibility.

We also remark that the $A$-terms are proportional to the Yukawa couplings, namely, $A_l=\frac{1}{2}(\gamma_{H_d}+\gamma_{{\tilde l}_L}+\gamma_{{\tilde e}_R})y_l  F_\Phi$ for sleptons. Thus, depending on the hierarchy between squark and slepton masses, the $A$-terms for sleptons can be sizable.

\subsection{Gauge coupling unification with split sparticle masses}

 In the presence of split sparticle masses in the MSSM and messenger sectors, it is important to check the quality of the unification of gauge couplings at the GUT scale $M_{\rm GUT}$. 
 
 Keeping in mind the general gauge mediation and assuming that the colored scalar superpartners are much heavier than the rest of the MSSM fields, we consider the running gauge couplings, as follows \cite{EOGM},
 \bea
 \alpha^{-1}_a (M_{\rm GUT})&=& \alpha^{-1}_a(M_Z) +\frac{b_{a,{\rm SM}}}{2\pi} \ln \frac{M_{\rm GUT}}{M_Z} \nonumber \\
 &&+\frac{b'_{a}}{2\pi} \ln \frac{M_{\rm GUT}}{m_{\rm SUSY}} +\frac{b_{a,{\tilde g}}}{2\pi} \ln \frac{M_{\rm GUT}}{m_{\tilde g}} +\frac{b_{a,{\tilde q}}}{2\pi} \ln \frac{M_{\rm GUT}}{m_{\tilde q}}-\frac{N}{2\pi} \ln \frac{M_{\rm GUT}}{\overline{{\cal M}_a}}
 \eea
 where $b_{a,{\rm SM}}=(-41/10, 19/6, 7)$ are the one-loop beta function coefficients in the SM,  $b'_a=(-7/5,-8/3,0)$ for the second Higgs doublet, electroweak gauginos, Higgsinos and sleptons, with common masses $m_{\rm SUSY}$,  $b_{a,{\tilde g}}=(0,0,-2)$ for gluinos,  $b_{a, {\tilde q}}=(-11/10,-3/2,-2)$ for  squarks with common masses $m_{\tilde q}$, and $ \overline{{\cal M}_a}$ are the effective mass scales for doublet and triplet messenger fields, given by
 \bea
 \overline{{\cal M}_a}=(X^n G(m_a,\lambda_a))^{1/N},  \quad a=2,3,
 \eea
with $G(m,\lambda)$ being some function of masses and couplings \cite{EOGM}, and $\overline{{\cal M}_{1}}=(\overline{{\cal M}_{2}})^{3/5}(\overline{{\cal M}_{3}})^{2/5}$ is the effective mass scale for hypercharged messenger fields. We note that $b_a=b_{a,{\rm SM}}+b'_{a}+b_{a,{\tilde g}}+b_{a, {\tilde q}}=(-33/5,-1,3)$ are the one-loop beta function coefficients in the MSSM.  In the absence of a doublet-triplet splitting for the messenger fields,   namely, $\overline{{\cal M}_{2}}=\overline{{\cal M}_{3}}$, the unified gauge couplings are achieved as in the MSSM.  
 
We now introduce the measure of unification as
\bea
B= \frac{\alpha^{-1}_2(M_Z) -\alpha^{-1}_3(M_Z) }{\alpha^{-1}_1(M_Z) -\alpha^{-1}_2(M_Z) }.
\eea
As compared to the MSSM where $B=\frac{b_3-b_2}{b_2-b_1}=\frac{5}{7}$ for $m_{\rm SUSY}=m_{\tilde g}=m_{\tilde q}\simeq M_Z$, we get a modified measure with the messenger fields for $m_{\rm SUSY}\simeq M_Z$ and $m_{\tilde g}, m_{\tilde q}$ being free, as
\bea
B=\frac{(b_3-b_2)\ln\big(\frac{M_{\rm GUT}}{M_Z}\big)+(b_{3,{\tilde g}}-b_{2,{\tilde g}})\ln\big(\frac{m_{\tilde g}}{M_Z}\big)+(b_{3,{\tilde q}}-b_{2,{\tilde q}})\ln\big(\frac{m_{\tilde q}}{M_Z}\big)+N\ln\big( \frac{\overline{{\cal M}_{3}}}{\overline{{\cal M}_{2}}}\big)}{(b_2-b_1)\ln\big(\frac{M_{\rm GUT}}{M_Z}\big)+(b_{2,{\tilde g}}-b_{1,{\tilde g}})\ln\big(\frac{m_{\tilde g}}{M_Z}\big)+(b_{2,{\tilde q}}-b_{1,{\tilde q}})\ln\big(\frac{m_{\tilde q}}{M_Z}\big)-\frac{2}{5}N\ln \big(\frac{\overline{{\cal M}_{3}}}{\overline{{\cal M}_{2}}}\big)}.
\eea
 Imposing the derivation from the MSSM to be no more than $5\%$, we get the bound on the extra differential running as
 \bea
 &&\bigg|\bigg(\frac{b_{3,{\tilde g}}-b_{2,{\tilde g}}}{b_3-b_2}-\frac{b_{2,{\tilde g}}-b_{1,{\tilde g}}}{b_2-b_1}\bigg)\ln\Big( \frac{m_{\tilde g}}{M_Z}\Big) + \bigg(\frac{b_{3,{\tilde q}}-b_{2,{\tilde q}}}{b_3-b_2}-\frac{b_{2,{\tilde q}}-b_{1,{\tilde q}}}{b_2-b_1}\bigg)\ln\Big( \frac{m_{\tilde q}}{M_Z}\Big) \\  &&\quad\quad+N\bigg(\frac{1}{b_3-b_2}+\frac{2}{5}\,\frac{1}{b_2-b_1}\bigg)\ln\frac{\overline{{\cal M}_{3}}}{\overline{{\cal M}_{2}}}  \bigg|\lesssim 0.036 \ln \bigg(\frac{M_{\rm GUT}}{M_Z}\bigg).
 \eea  
 Then, for $\ln(M_{\rm GUT}/M_Z)\simeq 33$, the quality of unification requires
 \bea
 \bigg|N\ln\frac{\overline{{\cal M}_{3}}}{\overline{{\cal M}_{2}}} -\frac{1}{2}\ln\Big( \frac{m_{\tilde g}}{M_Z}\Big) - \frac{1}{6}\ln\Big( \frac{m_{\tilde q}}{M_Z}\Big)\bigg|\lesssim 3.67. \label{unification}
 \eea
 Therefore, we find that there is a destructive interference between the split colored superpartners in the MSSM and the messenger fields with $ \overline{{\cal M}_{3}}\gtrsim  \overline{{\cal M}_{2}}$ in the differential running of the gauge couplings.
 
 Moreover, for $\overline{{\cal M}_{2}}\approx \overline{{\cal M}_{3}}\equiv \overline{{\cal M}}$, the perturbativity condition for the unified coupling gives rise to
 \bea
N\ln \frac{M_{\rm GUT}}{\overline{{\cal M}}}\lesssim 150. 
 \eea
 Thus, we can choose the number of messenger fields appropriately for perturbativity, for instance, $N=8,10,15, 20$ for $\overline{{\cal M}}=10^5 ,10^7, 10^9\,{\rm GeV}, 10^{13}\,{\rm GeV}$.

It was pointed out in Ref.~\cite{EOGM} that even if $\overline{{\cal M}_{2}}\approx \overline{{\cal M}_{3}}$, it is possible to realize an arbitrary amount of doublet-triplet splitting for the messenger fields, because $G(m,\lambda)$ appearing in the determinant of the mass matrix is generally independent of some of the couplings.  However, in our case, the unification condition in eq,~(\ref{unification}) can be achieved even for $ \overline{{\cal M}_{3}}\gtrsim  \overline{{\cal M}_{2}}$, in the presence of split colored superpartners in the MSSM.

We also remark that in the case of anomaly-gravity mixed mediation, there is no messenger sector which is charged under the SM gauge groups, so we only have to consider the MSSM fields for gauge coupling unification. Thus, setting the contribution of the messenger sector on the quality of unification in eq.~(\ref{unification}) to zero, we need the unification condition on the mass splitting in the MSSM sector, as follows,
\bea
\frac{1}{2}\ln\Big( \frac{m_{\tilde g}}{M_Z}\Big) + \frac{1}{6}\ln\Big( \frac{m_{\tilde q}}{M_Z}\Big)\lesssim 3.67.
\eea
Then, for $m_{\tilde g}\simeq 2\,{\rm TeV}$, we get the upper bound on the squark masses as $m_{\tilde q}\lesssim 3.14\times 10^7\,{\rm GeV}$.
This result is also true of the case in the general gauge mediation with $\overline{{\cal M}_{2}}\approx \overline{{\cal M}_{3}}$.
 
It is remarkable that the scalar soft masses below the messenger scale are subject to the renormalization group (RG) running due to gaugino masses. Namely, we need to include the RG running effects by $K_3+K_2+\frac{1}{36}K_1$, $K_3+\frac{4}{9}K_1$, $K_3+\frac{1}{9} K_1$ for $m^2_{{\tilde q}_L}$, $m^2_{{\tilde u}_R}$ and $m^2_{{\tilde d}_R}$, respectively,  $K_2+\frac{1}{4}K_1$, $K_1$ for $m^2_{{\tilde l}_L} $ and $m^2_{{\tilde e}_R}$, respectively. Here, $K_1=\frac{3}{10\pi^2}\int^{\ln \overline {{\cal M}_1}}_{\ln\mu} dt\, g^2_1(t) |M_1(t)|^2$, $K_2=\frac{3}{8\pi^2}\int^{\ln \overline {{\cal M}_2}}_{\ln\mu} dt\, g^2_2(t) |M_2(t)|^2$ and $K_3=\frac{2}{3\pi^2}\int^{\ln \overline {{\cal M}_3}}_{\ln\mu} dt\, g^2_3(t) |M_3(t)|^2$\cite{prime}. Then, for $N=N_2=10-20$ and $N_3={\cal O}(1)$ in general gauge mediation, the loop corrections are approximately given by $K_1\simeq (0.06-0.12) M^2_1(\overline{{\cal M}})\simeq (0.08-0.22)M^2_1$, $K_2\simeq (0.14-0.29) M^2_2(\overline{{\cal M}})=(0.15-0.34)M^2_2$ and $K_3\simeq (0.98-1.3)M^2_3(\overline{{\cal M}})\simeq (0.28-0.64)M^2_3$ where $M_a(\overline{{\cal M}})$ are the gaugino masses at the messenger scale, $\overline{{\cal M}_{2}}\approx \overline{{\cal M}_{3}}\equiv \overline{{\cal M}}$, and $M_a$ simply denote the gaugino masses at low energy. Since the RG running is small and squarks are relatively heavy, the loop corrections to the scalar soft masses are relatively small, as compared to the case where the messenger scale is the GUT scale: $K_1\simeq 0.15M^2_1(M_{\rm GUT})$, $K_2\simeq 0.5M^2_2(M_{\rm GUT})$ and $K_1\simeq (4.5-6.5)M^2_3(M_{\rm GUT})$  where $M_a(M_{\rm GUT})$ are the gaugino masses at the GUT scale \cite{prime}. Therefore, in general gauge mediation with split sparticles, it is sufficient to include the tree-level effects for the scalar soft masses due to electroweak symmetry breaking but ignore the loop corrections.

\section{Electroweak precision and vacuum stability bounds from sleptons}

In this section, we include the bounds from electroweak precision measurements such as the $W$ boson mass and the vacuum stability bounds on the $\mu$-term as complementary probes to test the models in addition to the muon $g-2$ constraints.

\subsection{Electroweak precision and $W$ boson mass}

For a large $\mu$-term, we can have a large slepton mixing so the contribution of the smuon loops to the muon $g-2$ can be enhanced.
However, the larger the $\mu$-term, the larger the mass splitting within the $SU(2)$ doublet slepton, causing a larger deviation in the electroweak precision data.

The theoretical value of the $W$ boson mass can be derived from the muon decay amplitude, which relates $M_W$ to the Fermi constant $G_\mu$, the fine structure constant $\alpha$, and the $Z$ boson mass $M_Z$, with the following modified formula,
\bea
M^2_W \bigg( 1-\frac{M^2_W}{M^2_Z}\bigg) =\frac{\pi \alpha}{\sqrt{2} G_\mu}  ( 1+ \Delta r ) 
\eea
where $\Delta r$ encodes the loop corrections in the SM and the contributions from new physics.
We note that $\Delta r=0.0381$ in the SM, which leads to the SM prediction for the $W$ boson mass, as follows \cite{Haller:2018nnx,ParticleDataGroup:2020ssz}, 
\bea
M^{\rm SM}_W=80.357\,{\rm GeV}\pm 6\,{\rm MeV}. 
\eea
On the other hand, the world average for the measured $W$ boson mass in PDG \cite{ParticleDataGroup:2020ssz} is given by 
\bea
M^{\rm PDG}_W= 80.379\,{\rm GeV}\pm 12\,{\rm MeV}.
\eea
Thus, the SM prediction for the $W$ boson mass is consistent with the PDG value within $2\sigma$.
However, the Fermilab CDFII experiment  \cite{CDF:2022hxs} has recently measured the $W$ boson mass as
\bea
M^{\rm CDFII}_W= 80.4335\,{\rm GeV}\pm 9.4\,{\rm MeV}. \label{dw}
\eea
So, if confirmed,  the result could show a considerable deviation from the SM prediction at the level of $7.0\sigma$, calling for a new physics explanation. 

In the MSSM, soft SUSY masses are $SU(2)_L$ invariant, but  $SU(2)_L$ breaking mass terms split between the masses of the scalar superpartners within the same $SU(2)_L$ doublet, so there can be a sizable contribution to the $\rho$ parameter \cite{delrho}, as follows,
\bea
\Delta \rho=\frac{3G_\mu}{8\sqrt{2}\pi^2} \bigg[-\sin^2\theta_{\tilde\mu}\cos^2\theta_{\tilde\mu} F_0(m^2_{{\tilde\mu}_1},m^2_{{\tilde\mu}_2})
+\cos^2\theta_{\tilde\mu} F_0(m^2_{\tilde\nu},m^2_{{\tilde\mu}_2})+\sin^2\theta_{\tilde\mu} F_0(m^2_{\tilde\nu},m^2_{{\tilde\mu}_1})\bigg]
\eea
with
\bea
F_0(x,y)=x+y -\frac{2xy}{x-y}\ln \frac{x}{y}.
\eea
As the new physics contribution is related to the correction to the $\rho$ parameter by 
\bea
(\Delta r)_{\rm new} = -\frac{c^2_W}{s^2_W}\,\Delta\rho, \label{deltar}
\eea
we can turn the correction to the $\rho$ parameter into the correction to the W boson mass by
\bea
\Delta M_W \simeq \frac{1}{2} M_W\,\frac{c^2_W}{c^2_W-s^2_W}\, \Delta \rho. \label{Wmass}
\eea
The global fit in PDG constrains the slepton masses and mixing by $\Delta\rho=(3.8\pm 2.0)\times 10^{-4}$ \cite{ParticleDataGroup:2020ssz}.

As will be shown in the next section, we find that the CDFII results \cite{CDF:2022hxs} cannot be explained in the parameter space explaining the muon $g-2$, but our results are consistent with the PDG results within $2\sigma$ \cite{ParticleDataGroup:2020ssz}.

\subsection{Vacuum stability bounds}

When the charged sleptons have a sizable mixing due to a large $\mu$-term, we need to consider the bound from vacuum instability, because there can be a deep minimum violating charge in the scalar potential. There is a recent analysis of the vacuum stability for the charged sleptons in Ref.~\cite{VSB}.

For the vacuum stability in the presence of a large $\mu$-term, we take the scalar potential for the neutral Higgs scalar $H_d^0$ and charged sleptons, ${\tilde e}_L, {\tilde e}_R$,
\bea
V&=&m^2_{LL} |{\tilde e}_L|^2 +m^2_{RR} |{\tilde e}_R|^2 +m^2_{H_d} |H^0_d|^2 +\frac{1}{2} \bigg|\sum_{i}g_a \phi^*_i T^a\phi_i\bigg|^2 \nonumber \\
&&+ (y_l A_l H^0_d {\tilde e}_L {\tilde e}^*_R -y_l \mu_H  (H^0_d)^* {\tilde e}_L {\tilde e}^*_R  +{\rm h.c.}) 
\label{effpot}
\eea
where  the fourth term corresponds to the D-term potential with group generators $T^a$ running over $SU(2)_L\times U(1)_Y$ and $\phi_i$ being the Higgs doublet  and the sleptons. Then, taking the direction of maximizing the negative contribution of trilinear terms, $|H^0_d|=|{\tilde e}_L|=|{\tilde e}_R|\equiv \frac{1}{\sqrt{6}}\phi$, in eq.~(\ref{effpot}), we obtain the effective potential as follows \cite{Aterm},
\bea
V=\frac{1}{2} {\widetilde m}^2 \phi^2 -\frac{1}{3\sqrt{6}}\, A\phi^3 + \frac{1}{36} \lambda^2 \phi^4,
\eea
with $A\equiv y_l |A^*_l-\mu_H\tan\beta|$, and 
\bea
 {\widetilde m}^2 &=&\frac{1}{3}(m^2_{LL}+ m^2_{RR}+ m^2_{H_d,{\rm eff}}), \\
 \lambda &=& \frac{1}{\sqrt{2}} g'.
\eea
Here, $m^2_{H_d,{\rm eff}}=\mu^2_H+ m^2_{H_d}$ is the effective mass square for the down-type Higgs doublet. Then, for a vanishing trilinear soft mass, i.e. $A_l=0$, the absolute stability at the origin, $\phi=0$, requires
\bea
\frac{\sqrt{2}m_\mu}{v\cos\beta}\, \cdot|\mu_H|\tan\beta < \sqrt{3} \lambda\, {\widetilde m}. \label{absolute}
\eea
On the other hand, the metastability of the false vacuum at the origin needs the the decay rate to be smaller than unity, namely,
\bea
(\Gamma/V) L^4 \ll 1,
\eea
where $L$ is the Hubble radius at present, $\Gamma/V\simeq {\widetilde m}^4\, e^{-S_E}$ is the decay rate per volume with $S_E$ being the Euclidean action. Therefore, from the Euclidean action along the direction of $|H^0_d|=|{\tilde e}_L|=|{\tilde e}_R|\equiv \frac{1}{\sqrt{6}}\phi$, we obtain the metastability bound \cite{Aterm} as
\bea
S_E> 400+ 4\ln ({\widetilde m}/1\,{\rm TeV}).
\eea
In the thin-wall approximation with $\lambda^2 {\widetilde m}^2/A^2\to 1/3^-$, we get $S_E=\frac{9\pi^2}{2} \big(\frac{{\widetilde m}^2}{A^2}\big)\big(1-3\lambda^2\,\frac{{\widetilde m}^2}{A^2}\big)^{-3}$  \cite{Aterm} . On the other hand, in the thick-wall limit with $|\lambda^2 {\widetilde m}^2/A^2|\ll 1$, the Euclidean action is numerically approximated to $S_E=1225\, {\widetilde m}^2/A^2$  \cite{Aterm}. 
In this case, for ${\widetilde m}\sim 1\,{\rm TeV}$ and $A_l=0$, the metastability bound becomes
\bea
\frac{\sqrt{2}m_\mu}{v\cos\beta} \, \cdot|\mu_H|\tan\beta\lesssim 1.75 {\widetilde m}, \label{metastab}
\eea
for $|\lambda^2|\ll 1$.

As a consequence, the $\mu$-term is bounded from the above, due to eq.~(\ref{absolute}) for absolute stability or eq.~(\ref{metastab}) for metastability. For instance, for ${\widetilde m}=1\,{\rm TeV}$, the absolute stability bound allows for the $\mu$-parameter up to $|\mu_H|_{\rm max}=1.1\,{\rm TeV}, 9.9\,{\rm TeV}$ for $\tan\beta=30, 10$, respectively, and the metastability bound in eq.~(\ref{metastab}) sets the maximum value of the $\mu$-parameter to $|\mu_H|_{\rm max}=3.2\,{\rm TeV}, 29\,{\rm TeV}$ for $\tan\beta=30, 10$, respectively.  If the ${\widetilde m}$ is larger than $1\,{\rm TeV}$, the $\mu$-parameter can be larger, for instance, due to a large soft mass for the down-type Higgs doublet. However, we take the vacuum stability bounds for ${\widetilde m}=1\,{\rm TeV}$ implicitly in the later discussion. 

For comparison, the $\mu$-term is also bounded by the slepton mixing angle for a given slepton mass splitting. When the $A_l$ term vanishes, eq.~(\ref{sleptonmix}) gives rise to a bound on the $\mu$-parameter as $ 2 m_l |\mu_H|\tan\beta\leq m^2_{{\tilde e}_2}-m^2_{{\tilde e}_1}$. Then, we also need to take the slepton mixing angle into account for choosing the $\mu$-parameter, together with the vacuum stability bounds. For instance, for $\sqrt{m^2_{{\tilde e}_2}- m^2_{{\tilde e}_1}}=100 (300)\,{\rm GeV}$, the $\mu$-parameter is bounded by $|\mu_H|_{\rm max}=1.6 (14)\,{\rm TeV}, 4.7 (42)\,{\rm TeV}$ for $\tan\beta=30, 10$, respectively. Therefore, for small slepton masses, the bounds from the slepton mixing angle are comparable to those from the vacuum stability.

In the later discussion on the muon $g-2$, we take into account the bounds on the $\mu$-term coming from the vacuum stability and the physical slepton masses.

\section{Muon $g-2$ from the sleptons}

We present the general formulas for the one-loop contributions of new charged or neutral particles to the muon $g-2$ and apply them to the case with smuons, sneutrino, electroweak gauginos and Higgsinos in MSSM. We also show how the muon $g-2$ anomaly can be explained with sparticle masses in general gauge mediation. We discuss the mass ordering of sleptons and the lightest neutralino, going beyond the scenarios where the lightest neutralino is the LSP.

\subsection{Muon $g-2$ from new scalars or fermions}

We list the general formulas for the muon $g-2$ due to new interactions to the muon.
First, suppose that there is an extra Yukawa type interaction of the muon to a neutral scalar $\phi$ with mass $m_\phi$ and a charged fermion $F$ with charge $-1$ and mass $m_F$, as follows,
\bea
{\cal L}_1 = - {\bar \mu} \Big(A_S + A_P \gamma^5 \Big)F\phi +{\rm h.c.}. \label{gen1}
\eea
Then,  the one-loop contribution to the muon $g-2$ is given \cite{leveille} by
\bea
a^{(1)}_\mu &=&- \frac{m^2_\mu}{8\pi^2}\int^1_0 dx\, \frac{|A_S|^2 \big(x^3-x^2-x^2\,\frac{m_F}{m_\mu}\big)+|A_P|^2 (m_F\rightarrow -m_F)}{(1-x)m^2_\phi+(m^2_F-m^2_\mu)x+x^2 m^2_\mu} \nonumber \\
&=& \frac{m_\mu}{8\pi^2} \bigg(\frac{m_\mu}{12m^2_\phi} (|A_S|^2+|A_P|^2) F^C_1(x_F)+\frac{m_F}{3m^2_\phi} (|A_S|^2-|A_P|^2) F^C_2(x_F)\bigg), \label{a1}
\eea
with $x_F=m^2_F/m^2_\phi$ and
\bea
F^C_1(x)&=& \frac{2}{(1-x)^4} \Big(2+3x-6x^2+x^3+6x\ln x\Big),  \\
F^C_2(x)&=&-\frac{3}{2(1-x)^3} \Big(3-4x+x^2+2\ln x\Big).
\eea

Similarly, in the presence of an extra Yukawa type interaction of the muon to a charged scalar $\chi$ with charge $-1$ and mass $m_\chi$ and a neutral fermion $\lambda$ with mass $m_\lambda$, given by
\bea 
{\cal L}_2 = -{\bar \mu} \Big(C_S + C_P \gamma^5 \Big)\lambda\chi +{\rm h.c.}, \label{gen2}
\eea
we obtain the one-loop contribution to the muon $g-2$  \cite{leveille} as
\bea
a^{(2)}_\mu &=& \frac{m^2_\mu}{8\pi^2}\int^1_0 dx\, \frac{|C_S|^2 \big(x^3-x^2+\frac{m_\lambda}{m_\mu}(x^2-x)\big)+|C_P|^2 (m_\lambda\rightarrow -m_\lambda)}{(1-x)m^2_\lambda+(m^2_\chi-m^2_\mu)x+x^2 m^2_\mu} \nonumber \\
&=& \frac{m_\mu}{8\pi^2}\bigg(-\frac{m_\mu}{12m^2_\chi} (|C_S|^2+|C_P|^2 )F^N_1(x_\lambda)-\frac{m_\lambda}{6m^2_\chi} (|C_S|^2-|C_P|^2)F^N_2(x_\lambda)  \bigg), \label{a2}
\eea
with $x_\lambda=m^2_\lambda/m^2_\chi$ and
\bea
 F^N_1(x) &=& \frac{2}{(1-x)^4} \Big(1-6x+3x^2+2x^3-6x^2\ln x\Big), \\
 F^N_2(x) &=&\frac{3}{(1-x)^3} \Big( 1-x^2+2x\ln x\Big).
\eea

\begin{figure}[!t]
\centering
\includegraphics[width=0.70\textwidth,clip]{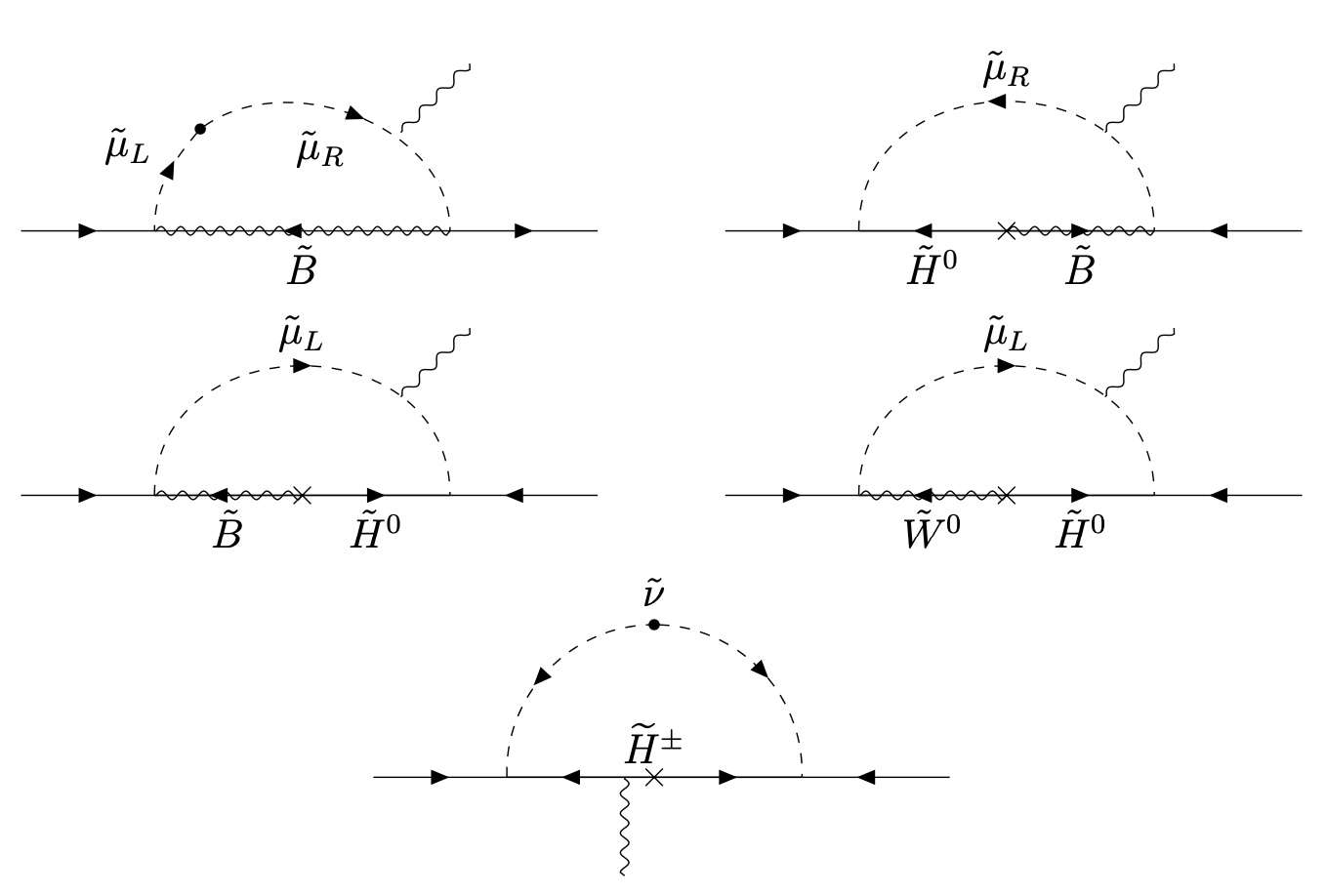}
\caption{ Feynman diagrams with smuons and sneutrino, contributing to the muon $g-2$.
}
\label{g2feynman}
\end{figure}

\subsection{Muon $g-2$ from sleptons}

Now we are in a position to apply the general formulas for the muon $g-2$ to the supersymmetric case with sleptons \cite{g2mssm,g2-recent}. Feynman diagrams with sleptons relevant for the muon $g-2$ are shown in Fig.~\ref{g2feynman}.

In the basis of mass eigenstates for neutralinos, charginos and sleptons, given in eqs.~(\ref{neutralinomix}), (\ref{charginomix}) and (\ref{sleptonmix}),
we rewrite the the slepton interactions relevant for the muon $g-2$ in eq.~(\ref{sleptonint}), as follows,
\bea
{\cal L}_{\rm sleptons} =-\sqrt{2} g' {\bar \mu}\Big(B^{L}_{i,a} P_L + B^{R}_{i,a} P_R\Big){\widetilde\chi}^0_i {\tilde \mu}_a  -{\tilde\nu}_{\mu L} {\bar\mu} \Big( g\, U^*_{R,i1} P_R -f_\mu U^*_{L,i2}P_L\Big){\widetilde\chi}^-_i +{\rm h.c.}  \label{effsleptons}
\eea
with
\bea
B^{L}_{i,1} &=& \cos\theta_{\tilde\mu}Y_{e^c_R} N^*_{i1}+\frac{1}{\sqrt{2}g'}\, f_\mu\sin\theta_{\tilde\mu} \, N^*_{i3}, \\
B^{L}_{i,2} &=& \sin\theta_{\tilde\mu}Y_{e^c_R} N^*_{i1}+\frac{1}{\sqrt{2}g'}\, f_\mu\cos\theta_{\tilde\mu} \, N^*_{i3}, \\
B^{R}_{i,1} &=& -\sin\theta_{\tilde\mu} \Big(Y_{l_L} N_{i1}+\cot\theta_W\, T^3(\mu_L) N_{i2} \Big) -\frac{1}{\sqrt{2}g'}\, f_\mu\cos\theta_{\tilde\mu}\, N_{i3}, \\
B^{R}_{i,2} &=& \cos\theta_{\tilde\mu} \Big(Y_{l_L} N_{i1}+\cot\theta_W\, T^3(\mu_L) N_{i2} \Big) -\frac{1}{\sqrt{2}g'}\, f_\mu\sin\theta_{\tilde\mu}\, N_{i3}.
\eea
Here, $f_\mu$ is the muon Yukawa coupling, given by $f_\mu=\sqrt{2}m_\mu/(v\cos\beta)$.  
As compared to the general muon Yukawa interactions in eqs.~(\ref{gen1}) and (\ref{gen2}), we can match the slepton interactions by
\bea
{\cal L}_{\rm sleptons}=-{\bar\mu} \Big( C^{(i,a)}_{S}+C^{(i,a)}_{P} \gamma^5 \Big){\widetilde\chi}^0_i {\tilde \mu}_a -{\tilde\nu}_{\mu L} {\bar\mu}\Big(A^{(j)}_S+A^{(j)}_P \gamma^5\Big) {\widetilde\chi}^-_j  +{\rm h.c.} \label{effint}
\eea
with
\bea
C^{(i,a)}_{S} &=&  \frac{g'}{\sqrt{2}}\, \Big( B^{R}_{i,a} +B^{L}_{i,a} \Big), \quad a=1,2,  \\
C^{(i,a)}_{P} &=&\frac{g'}{\sqrt{2}}\, \Big( B^{R}_{i,a} -B^{L}_{i,a} \Big),   \quad a=1,2,
\eea
and
\bea
A^{(j)}_S &=&  \frac{1}{2} \Big(g\, U^*_{R,j1}- f_\mu U^*_{L,j2} \Big), \\
A^{(j)}_P&=&   \frac{1}{2} \Big(g\, U^*_{R,j1}+ f_\mu U^*_{L,j2} \Big).
\eea
Then, we get
\bea
|C^{(i,a)}_{S}|^2+|C^{(i,a)}_{P}|^2&=&g^{\prime 2} \Big(|B^{R}_{i,a} |^2+|B^{L}_{i,a} |^2 \Big), \\
|C^{(i,a)}_{S}|^2-|C^{(i,a)}_{P}|^2&=& g^{\prime 2}\Big(B^{R}_{i,a}B^{L*}_{i,a}+B^{R*}_{i,a}B^{L}_{i,a}\Big), \\
|A^{(j)}_S|^2 +|A^{(j)}_P |^2 &=&\frac{1}{2} \Big(g^2 |U^*_{R,j1}|^2+f^2_\mu |U^*_{L,j2}|^2 \Big), \\
|A^{(j)}_S|^2 -|A^{(j)}_P |^2 &=&-\frac{1}{2}g f_\mu \Big(U^*_{R,j1}U_{L,j2}+U_{R,j1}U^*_{L,j2}\Big).
\eea

\begin{figure}[!t]
\centering
\includegraphics[width=0.45\textwidth,clip]{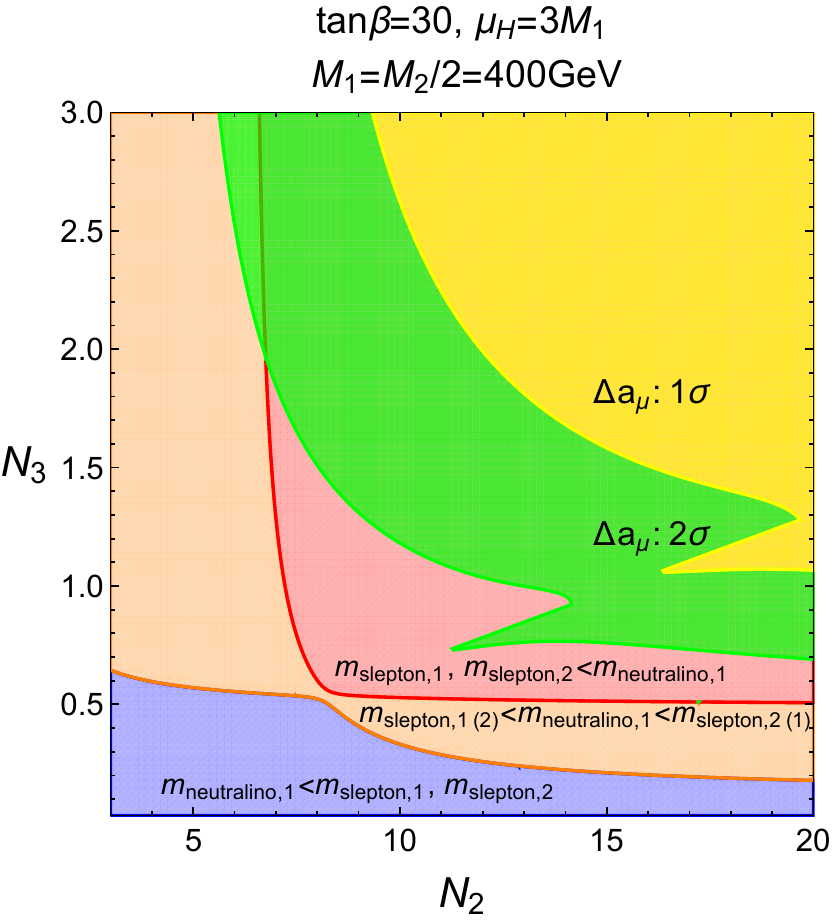}\,\,\,\,\,\,
\includegraphics[width=0.45\textwidth,clip]{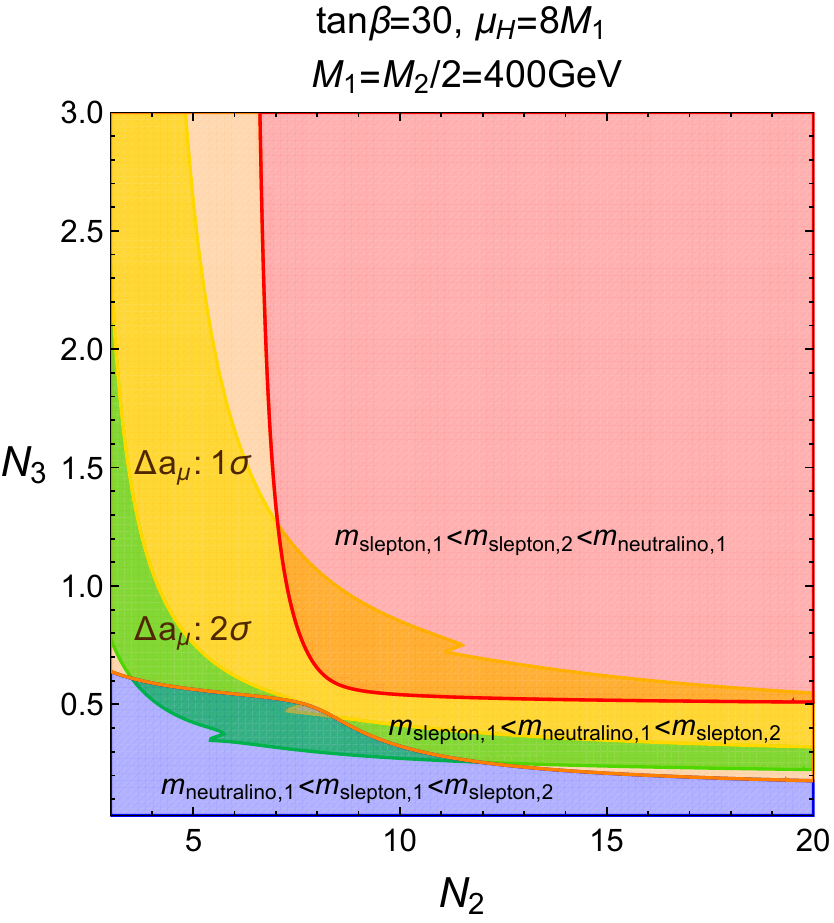}
\caption{Parameter space for $N_2$ and $N_3$, explaining the muon $g-2$. We indicated the mass ordering of smuons and the lightest neutralino. We took $\tan\beta=30$, $M_1=M_2/2=400\,{\rm GeV}$ for both plots, and $\mu_H=3M_1$ on left and $\mu_H=8M_1$ on right.
}
\label{fig:g-2a}
\end{figure}

As a result, using eqs.~(\ref{a1}) and (\ref{a2}), we obtain the full one-loop corrections to the muon $g-2$ due to the slepton loops \cite{g2mssm} as
\bea
\Delta a_\mu =a^{(1)}_\mu +a^{(2)}_\mu
\eea
with
\bea
a^{(1)}_\mu &=&\sum_{j=1}^2\bigg[\frac{m^2_\mu}{96\pi^2 m^2_{\tilde\nu}} F^C_1(m^2_{{\tilde\chi}^-_j}/m^2_{\tilde\nu})\Big(g^2 |U^*_{R,j1}|^2+f^2_\mu |U^*_{L,j2}|^2 \Big) \nonumber \\
&&-\frac{g f_\mu m_\mu m_{{\tilde\chi}^-_j}}{48\pi^2m^2_{\tilde\nu}} F^C_2(m^2_{{\tilde\chi}^-_j}/m^2_{\tilde\nu}) \Big(U^*_{R,j1}U_{L,j2}+U_{R,j1}U^*_{L,j2}\Big)\bigg], \\
a^{(2)}_\mu &=&\sum_{a=1}^2\sum_{i=1}^4\bigg[-\frac{g^{\prime 2}m^2_\mu}{192\pi^2m^2_{{\tilde\mu}^2_a}} F^N_1(m^2_{{\tilde\chi}^0_i}/m^2_{{\tilde\mu}_a})\Big(|B^{R}_{i,a} |^2+|B^{L}_{i,a} |^2 \Big) \nonumber \\
&&-\frac{ g^{\prime 2} m_\mu m_{{\tilde\chi}^0_i}}{48\pi^2 m^2_{{\tilde\mu}^2_a}} F^N_2(m^2_{{\tilde\chi}^0_i}/m^2_{{\tilde\mu}_a}) \Big(B^{R}_{i,a}B^{L*}_{i,a}+B^{R*}_{i,a}B^{L}_{i,a}\Big)\bigg].
\eea
Here, the muon Yukawa coupling is given by $f_\mu=\sqrt{2} m_\mu/(v\cos\beta)$, so the terms with the muon Yukawa coupling can be enhanced for a large $\tan\beta$.

\begin{figure}[!t]
\centering
\includegraphics[width=0.45\textwidth,clip]{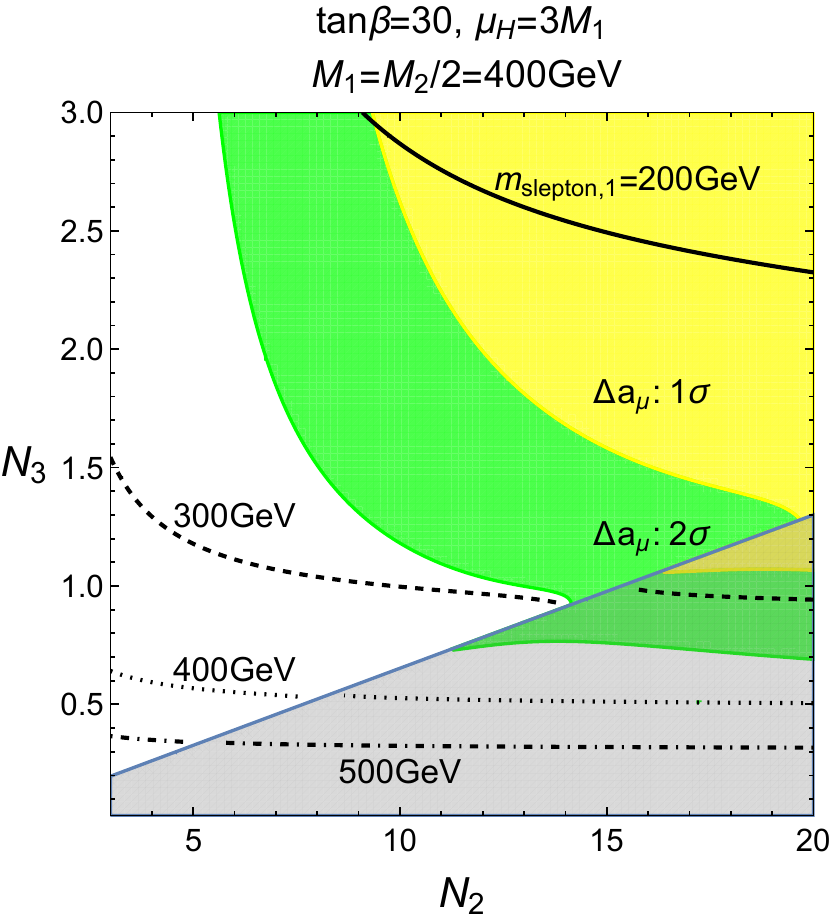}\,\,\,\,\,\,
\includegraphics[width=0.45\textwidth,clip]{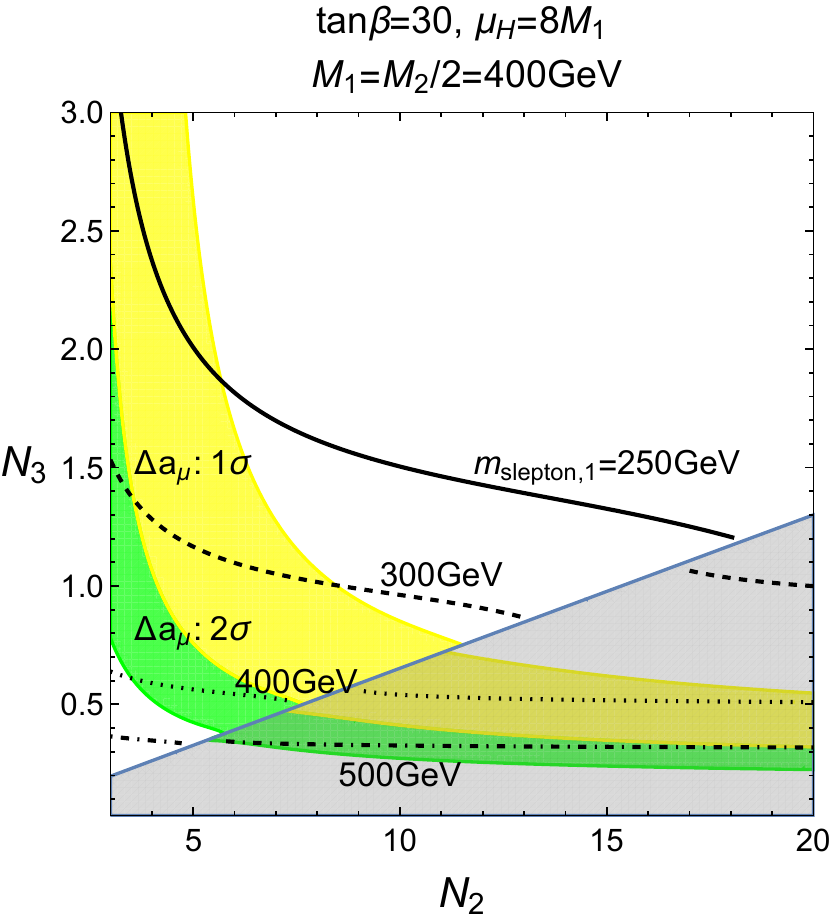}
\caption{The same for the muon $g-2$ constraints as in Fig.~\ref{fig:g-2a}, except that we also showed the contours for the lighter smuon mass, $m_{{\tilde\mu}_1}=200(250) {\rm GeV}, 300\,{\rm GeV}, 400\,{\rm GeV}, 500\,{\rm GeV}$ in solid, dashed, dotted and dot-dashed lines, respectively. We also showed the gray region where $m_{{\tilde\mu}_2}<m_{{\tilde\mu}_1}$ or $m^2_{LL}<m^2_{RR}$ so the lighter smuon is $SU(2)_L$ doublet-like. 
}
\label{fig:g-2b}
\end{figure}

For the contributions from sneutrino and charginos, we can approximate the one-loop corrections to the muon $g-2$ as
\bea
a^{(1)}_\mu&\simeq&  \frac{g^2 \,m_\mu^2 m_{{\tilde\chi}^-_2} }{24\pi^2 (M_2^2-\mu_H^2)m_{\tilde\nu}^2} (\mu_H+M_2 \tan\beta) F^C_2(m^2_{{\tilde\chi}^-_2}/m^2_{\tilde\nu}) \nonumber \\
&&- \frac{g^2\, m_\mu^2 m_{{\tilde\chi}^-_1} }{24\pi^2(M_2^2-\mu_H^2)m_{\tilde\nu}^2} (\mu_H \tan\beta+M_2) F^C_2(m^2_{{\tilde\chi}^-_1}/m^2_{\tilde\nu}). 
\eea
Here, we note that $xF^C_2(x)$ is positive and monotonically increasing for $x>0$.
Thus, for a large $\tan\beta$, we find that the chargino contribution for the muon $g-2$ can be positive.

Similarly, the contributions from smuons and bino-like neutralino can be enhanced for a large $\tan\beta$ or a sizable slepton mixing. In this case, in the limit of neglecting the effects of electroweak symmetry breaking, we also obtain the corresponding one-loop corrections to the muon $g-2$ approximately as
\bea
a^{(2)}_\mu\simeq \frac{g^{\prime 2}m_\mu m_{{\tilde\chi}^0_1} }{96\pi^2}\,\sin2\theta_{\tilde\mu} \bigg(\frac{1}{m^2_{{\tilde\mu}_2}} F^N_2(m^2_{{\tilde\chi}^0_1}/m^2_{{\tilde\mu}_2})-\frac{1}{m^2_{{\tilde\mu}_1}}F^N_2(m^2_{{\tilde\chi}^0_1}/m^2_{{\tilde\mu}_1})\bigg).
\eea
Here, we note that $xF^N_2(x)$ is positive and monotonically increasing for $x>0$. Thus, for  $\mu_H>0$, we can get a positive contribution to the muon $g-2$ for either $\sin2\theta_{\tilde\mu} <0$ (or $m^2_{{\tilde\mu}_2}>m^2_{{\tilde\mu}_1}$) (for which the lighter smuon is $SU(2)_L$ singlet-like) or $\sin2\theta_{\tilde\mu} >0$ (or $m^2_{{\tilde\mu}_2}<m^2_{{\tilde\mu}_1}$) (for which the lighter smuon is $SU(2)_L$ doublet-like). 

In Fig.~\ref{fig:g-2a}, we show the parameter space for $N_2$ and $N_3$ in general gauge mediation, explaining the muon $g-2$ within $1\sigma$ and $2\sigma$ in yellow and green regions, respectively.  We chose $\tan\beta=30$, $M_1=M_2/2=400\,{\rm GeV}$ for both plots, and $\mu_H=3M_1$ on left and $\mu_H=8M_1$ on right. We took $M_1=400\,{\rm GeV}$ to make the gluino mass, $M_3\simeq 6M_1$, sufficiently large for the LHC bounds \cite{gluino}, and  the chosen values of the $\mu$-parameter are consistent with the bounds from the vacuum stability in eqs.~(\ref{absolute}) or (\ref{metastab}).  In both plots, we divide the parameter space depending on the mass ordering of smuons and the lightest neutralino: $m_{{\tilde\mu}_1}<m_{{\tilde\mu}_2}<m_{{\tilde\chi}^0_1}$, $m_{{\tilde\mu}_1}<m_{{\tilde\chi}^0_1}<m_{{\tilde\mu}_2}$, $m_{{\tilde\chi}^0_1}<m_{{\tilde\mu}_1}<m_{{\tilde\mu}_2}$, in red, orange and blue regions, respectively.  For this example, the lighter smuon mass between $200\,{\rm GeV}$ and $300\,{\rm GeV}$ is consistent with the muon $g-2$ within $2\sigma$, but the lighter slepton is lighter than the lightest neutralino. 

In Fig.~\ref{fig:g-2b}, we show the parameter space for $N_2$ and $N_3$ with the same set of other parameters as in Fig.~\ref{fig:g-2a}, but we add the bounds from the $W$ boson mass in both plots and choose a larger $\mu$-parameter, $\mu_H=8M_1$ on right. In the right(left) plot of Fig.~\ref{fig:g-2a}, we also indicated the contours of the lighter charged slepton mass,  $m_{{\tilde\mu}_1}=200(250) {\rm GeV}, 300\,{\rm GeV}, 400\,{\rm GeV}, 500\,{\rm GeV}$, by solid, dashed, dotted and dot-dashed lines, respectively. We also showed the parameter space with $m_{{\tilde\mu}_2}<m_{{\tilde\mu}_1}$ or $m^2_{LL}<m^2_{RR}$ in gray where the lighter smuon is $SU(2)_L$ doublet-like.  We note that the entire parameter space above the orange region is consistent with $\Delta\rho$ bounds  in PDG at $2\sigma$, but the $W$ boson mass measured by CDFII cannot be explained.

From the right plot with $\mu_H=8M_1$ in Fig.~\ref{fig:g-2b}, we also show that there is a parameter space with the lighter smuon being heavier than the lightest neutralino, explaining the muon $g-2$ anomaly within $2\sigma$  for $N_3\lesssim 0.6$ and $N_2\lesssim12$. In most of the parameter space in $N_2$ and $N_3$ for the muon $g-2$, the lighter smuon is lighter than the lightest neutralino, which requires going beyond the standard searches for sleptons.

\begin{table}[!t]
    \centering
    \begin{tabular}{|c|c|c|c|}
    \hline
         & Model I & Model II  & Model III \\
        \hline
        $(N_2,N_3)$ & $(7,0.4)$ & $(20,1.5)$ & (11, 0.5) \\
        \hline
        $m_{{\tilde q}_L}$  &   6968  &  3600 & 6230 \\
        \hline 
        $m_{{\tilde u}_R}$   &   6964     &  3596 & 6229  \\
        \hline
        $m_{{\tilde d}_R}$     &  6959   &  3594   & 6224 \\
        \hline
        $m_{{\tilde l}_L}$    &   446  &   256  &   367   \\
        \hline
         $m_{{\tilde e}_R}$    &  467  &   244   & 414    \\
        \hline 
        $M_1$ &  409  &     409 & 409    \\
        \hline
        $M_2$  & 820    &    820    & 820   \\
        \hline
        $M_3$ &  2694    &     2694    & 2694       \\
        \hline
        $\mu_H$  &  3271 &   1227 &  3271  \\
        \hline
         $\tan\beta$  &  30 &  30 & 30  \\
        \hline
    \end{tabular}
    \\ \vspace{0.5cm}
      \begin{tabular}{|c|c|c|c|}
    \hline
          & Model I  & Model II   & Model III \\
        \hline
        $m_{\tilde \nu}$ &  441 &    247 & 361 \\
        \hline
        $m_{{\tilde\mu}_1}$ & 474  &   244  & 420    \\
        \hline
         $m_{{\tilde\mu}_2}$ & 443   &      264  & 366        \\
        \hline
         $m_{{\tilde\chi}^0_1}$  &  409 &   408  & 409    \\
        \hline 
         $m_{{\tilde\chi}^0_2}$ & 820  &     813 & 820       \\
        \hline
          $m_{{\tilde\chi}^0_3}$  &  3273 &  1236    & 3273     \\
        \hline
          $m_{{\tilde\chi}^0_4}$ & 3270 &   1225   & 3270           \\
        \hline
           $m_{{\tilde\chi}^\pm_1}$ &  820  &   813 & 820          \\
        \hline
         $m_{{\tilde\chi}^\pm_2}$ &  3273 & 1236  & 3273  \\
        \hline
          $\Delta a_\mu$ &  $1.56\times 10^{-9}$ &  $2.00\times 10^{-9}$  & $2.24\times 10^{-9}$  \\
        \hline
        $\Delta M_W$ &  $1.12\,{\rm MeV}$ &  $3.73\,{\rm MeV}$  & $1.73\,{\rm MeV}$ \\
        \hline
    \end{tabular}
    \caption{(Upper) Soft mass parameters in units of GeV for Model I, II and III in general gauge mediation. (Lower) Electroweak superpartner masses in units of GeV,  $\Delta a_\mu$ and $\Delta M_W$ for Model I, II and III. }
    \label{table:1}
\end{table}

In Table 1, we present three benchmark models which are consistent with the muon $g-2$ anomaly within $1\sigma$ (Model II and III) or $2\sigma$ (Model I).  We listed the input parameters for the benchmark models  in the upper table and the predicted masses for electroweak superpartners, $\Delta a_\mu$ and $\Delta M_W$ in each model in the lower table. In particular, Model III is also consistent with the $W$ boson mass measured by CDFII within $1\sigma$.  In the following, we discuss briefly the collider signatures for the benchmark models at the LHC.

In Model I, the sleptons are heavier than the lightest neutralino, so the lightest neutralino could be a dark matter candidate if there is no extra supersymmetric particle lighter than the lightest neutralino, such as gravitino. If the SUSY breaking in gauge mediation is the only source for the gravitino mass, the gravitino can be the LSP and a dark matter candidate, because the gravitino mass is given by $m_{3/2}=F/M_P$ and the Planck scale, $M_P$, is much greater than the messenger scale in gauge mediation \cite{gauge-review}. In this case, the lightest neutralino could decay into gravitino and the SM particle such as $Z$ or Higgs bosons. The lifetime of the LSP $X$ with mass $m_X$ in MSSM is given by $\tau_X=1.8\times 10^3\,{\rm sec}\,(m_{3/2}/100\,{\rm GeV})^2(1\,{\rm TeV}/m_X)^5$, which is constrained to be less than $5\times 10^3\,{\rm sec}$ by Big Bang Nucleosynthesis (BBN) \cite{BBN}.  If the lightest neutralino decays at the collider scale, the long-lived particle searches with displaced vertices at the LHC are applicable \cite{LLP}. Otherwise, we can consider the bounds from the standard SUSY searches with missing transverse momentum at the LHC \cite{sleptonsearch}.

In Model II and III, the sleptons are lighter than the lightest neutralino and the sneutrinos would become the LSP in the MSSM.  In this case, the heavier charged lepton (${\tilde\mu}_2$) can be produced at the LHC, decaying through a three-body decay mode, ${\tilde\mu}_2\to ({\tilde\chi}^0_1)^* \mu\to {\tilde\mu}_1 {\bar\mu}\mu$, where the lighter charged lepton (${\tilde\mu}_1$) decays in cascade through a three-body decay mode, ${\tilde\mu}_1\to ({\tilde\chi}^0_1)^* \mu\to {\tilde\nu} {\bar\nu}\mu$. The lighter charged lepton (${\tilde\mu}_1$)  can be produced at the LHC on its own, undergoing the same three-body decay process. If the sneutrinos are lighter than the gravitino, the sneutrino could be a dark matter candidate, but they should not be dominant components of dark matter, because it would have been already excluded by the direct detection bounds. Instead, if the sneutrinos are heavier than the gravitino, the gravitino can be a dark matter candidate, and the sneutrinos decay into gravitino and neutrino, being also constrained by BBN \cite{BBN}. In this case, the collider signatures are challenging, because multiple invisible particles are produced  promptly or non-promptly per each charged slepton in the final states.

\section{Proton decays from the sleptons}

We discuss the proton decay due to dimension-5 operators induced by colored Higgsinos in the minimal SU(5) GUT and its extension to orbifold GUT models in the extra dimensions.
We assume that squark and slepton masses are generation-independent but they can be split, as in general gauge mediation.

\subsection{Dimension-5 operators in the minimal SU(5)}

In the minimal SUSY SU(5) GUT, the Yukawa couplings between the MSSM matter fields, embedded in three copies of $\bf {\bar 5}+10$ representations, $\Psi^{ab}_i$ and $\Phi_{ja}$, with $a,b=1,2,\cdots, 5$ and $i,j=1,2,3$, and the MSSM Higgs fields, embedded in a pair of $5$ and ${\bar 5}$ representations, $H^a$ and ${\bar H}_a$, are given by
\bea
W_Y=\frac{1}{4} h^{ij} \epsilon_{abcde} \Psi^{ab}_i \Psi^{cd}_j H^e-\sqrt{2} f^{ij} \Psi^{ab}_i \Phi_{ja} {\bar H}_b. \label{dim5}
\eea
Here, the $\bf 5$ and $\bf {\bar 5}$ Higgs multiplets are $H^T=(H^\alpha_C, H^+_u, H^0_u)$ and ${\bar H}^T=({\bar H}_{C\alpha}, H^-_d,-H^0_d)$ with $\alpha=1,2,3$, which contain not only the MSSM Higgs fields, $H_u^T=(H^+_u,H^0_u)$ and $H_d^T=(H^0_d,H^-_d)^T$, but also the colored Higgs fields, $H^\alpha_C$ and ${\bar H}_{C\alpha}$.

Then, after the colored Higgsinos are integrated out, we get the effective dimension-5 operators in the superpotential, as follows \cite{SU5},
\bea
W_5&=&\frac{\kappa}{2M_{H_C}}\, f_{u_i} f_{d_l} V^*_{kl} e^{i\varphi_i} \epsilon_{\alpha\beta\gamma}\epsilon_{rs} \epsilon_{tu} Q^{\alpha r}_i Q^{\beta s}_i Q^{\gamma t}_k L^u_l \nonumber \\
&&+ \frac{\kappa}{M_{H_C}}\, f_{u_i} e^{i\varphi_i} f_{d_l} V^*_{kl} \epsilon^{\alpha\beta\gamma} {\overline U}_{i\alpha} {\overline E}_{i} {\overline U}_{k\beta} {\overline D}_{l\gamma} \label{dim5}
\eea 
where the SU(5) Yukawa couplings for up-type and down-type fermions are parametrized by $h^{ij}=f_{u_i} e^{i\varphi_i} \delta_{ij}$ and $f^{ij} =V^*_{ij} f_{d_j}$ with $V_{ij}$ being the Cabibbo-Kobayashi-Maskawa(CKM) matrix. Here, we note that $\kappa$ parametrizes the suppression of the couplings of the colored Higgsinos as in orbifold GUT models \cite{orbifolds} where the boundary conditions for the colored Higgsinos in the extra dimensions suppress the couplings between the colored Higgsinos and the MSSM matter fields at the orbifold fixed points in the extra dimension. 

We elaborate more on a mechanism to suppress the Yukawa couplings of the colored Higgsinos. 
In the minimal 5D $SU(5)$ GUT compactified on $S^1/{Z_2\times Z'_2}$, we can assign parities for the $SU(5)$ gauge fields under the $Z_2$ parities such that  the $SU(5)$ GUT group is broken down to $SU(3)\times SU(2)\times U(1)$ \cite{orbifolds,orbi-dim5}. In this case, the Dirichlet boundary conditions can be imposed on the colored Higgsinos living in the bulk as the full $\bf 5$ and $\bf {\bar 5}$ representations, so the tree-level Yukawa couplings for the colored Higgsinos are forbidden at the orbifold fixed point of the extra dimension  where $SU(5)$ is broken \cite{orbi-dim5}.  However, we can also introduce higher dimensional operators for the Yukawa couplings  at the orbifold fixed points such as $\frac{1}{M^{n}_*}h^{ij} \epsilon_{abcde} \Psi^{ab}_i \Psi^{cd}_j (\partial_5)^n H^e$ and $ \frac{1}{M^{n}_*} f^{ij} \Psi^{ab}_i \Phi_{ja} (\partial_5)^n {\bar H}_b$  with $n$ being odd and $M_*$ being the fundamental scale in 5D. In this case, the suppression factor $\kappa$ in eq.~(\ref{dim5}) is obtained as $\kappa=1/(M_* R)^{n}$ with $R$ being the radius of the extra dimension, so we can obtain small Yukawa couplings for the colored Higgsinos as far as $M_*R \gg 1$. For instance, for $M_* R=20$ and $n=3$, we can achieve a suppression factor, $\kappa\sim 10^{-4}$. Such a suppression factor in the extra dimension is comparable to the case where the selection rules due to discrete $R$ symmetries respect the baryon number conservation in the leading order terms \cite{Rsym}.

\begin{figure}[t]
\centering
\includegraphics[width=0.60\textwidth,clip]{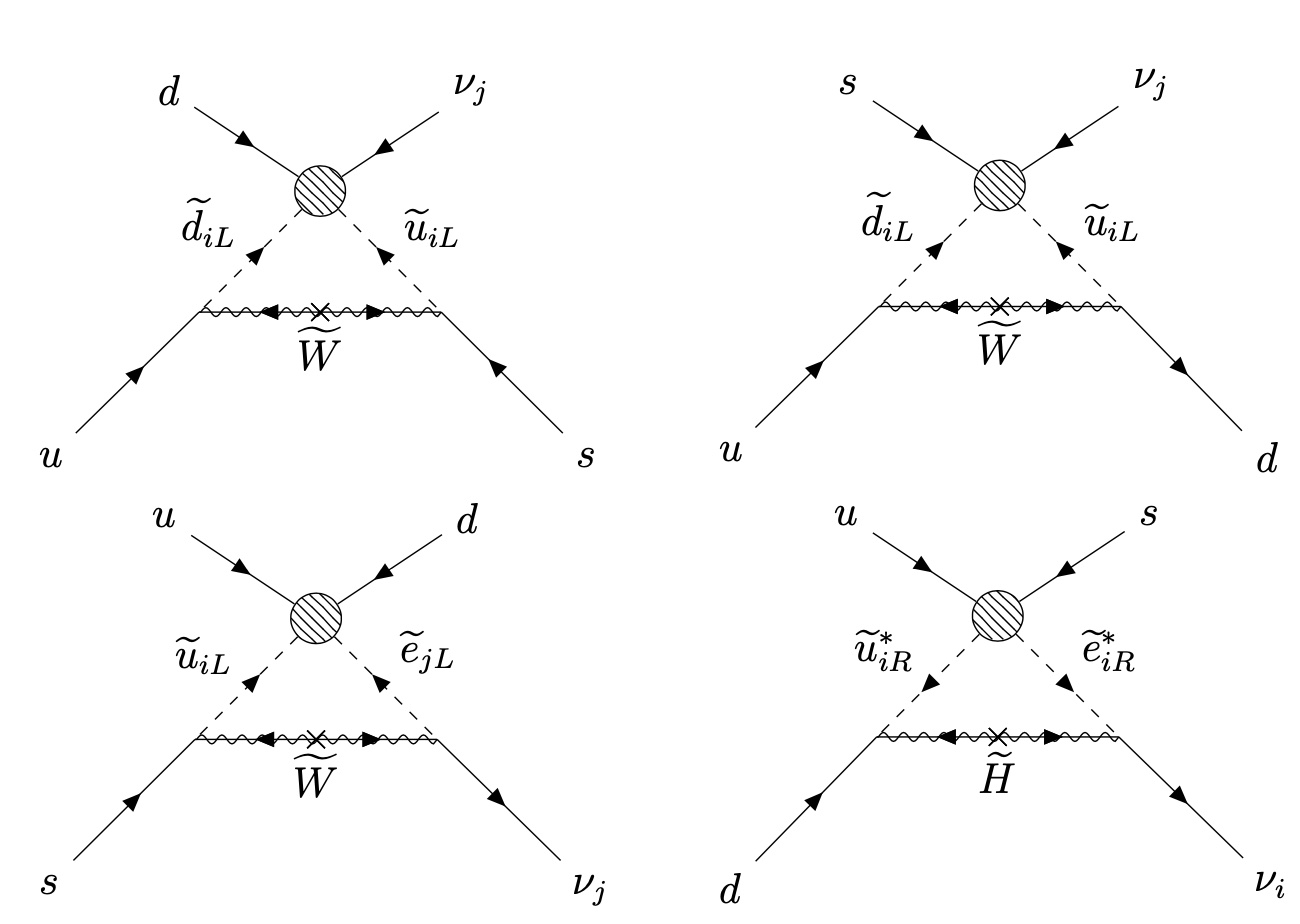}
\caption{Feynman diagrams with dimension-5 operators for proton decay mode, $p\to K^+{\bar\nu}$.
}
\label{protondecay}
\end{figure}

The above superpotential in eq.~(\ref{dim5}) gives rise to the effective dimension-5 interactions relevant for the proton decay between a pair of SM fermions and a pair of scalar superpartners in components,
\bea
{\cal L}_{{\rm dim-5}} &=& -\frac{\kappa}{M_{H_C}}\, f_{u_i} f_{d_l} V^*_{kl} e^{i\varphi_i} \bigg[(u_{iL} d_{iL}) ({\tilde u}_{kL} {\tilde e}_{lL})+({\tilde u}_{iL} d_{iL}) (u_{kL} {\tilde e}_{lL})- (u_{iL} d_{iL})({\tilde d}_k {\tilde \nu}_l)    \nonumber \\
&&\quad- (u_{iL} {\tilde d}_{iL})(d_k {\tilde \nu}_l) - (u_{iL} {\tilde d}_{iL})({\tilde d}_k \nu_l) - ({\tilde u}_{iL} d_{iL})({\tilde d}_k \nu_l) - ({\tilde u}_{iL} {\tilde d}_{iL})(d_k \nu_l)\bigg] \nonumber \\
&&- \frac{\kappa}{M_{H_C}}\, f_{u_i} e^{i\varphi_i} f_{d_l} V^*_{kl}  \bigg[(u^c_{iR}{\tilde e}^*_{iR} {\tilde u}^*_{kR}d^c_{lR} )+({\tilde u}^*_{iR}{\tilde e}^*_{iR} u^c_{kR}d^c_{lR} ) \bigg].
\eea
Thus, one-loop corrections with squarks and/or sleptons give rise to the proton decay mode,  $p\to K^+{\bar\nu}$, as shown in Fig.~\ref{protondecay}.
Therefore, the resultant lifetime of the proton depends on squark and slepton masses as well as chargino masses.

\subsection{Proton decays with split sparticle masses}

Taking into account one-loop diagrams in Fig.~\ref{protondecay}, we get the effective dimension-6 operators for proton decay, $p\to K^+ {\bar \nu}_j$, with $j=2,3$, as follows \cite{SU5},
\bea
{\cal L}_{\rm dim-6} &=& \frac{\kappa\alpha^2_2}{M_{H_C}m^2_W \sin2\beta } \bigg[\sum_{i,j=2,3} 2F(M_2, m^2_{{\tilde d}_{iL}},m^2_{{\tilde u}_{iL}}) \,{\overline m}_{u_i}{\overline m}_{d_j}  V_{u_i d} V_{u_i s} V^*_{ud_j} e^{i\varphi_i}  \nonumber \\
&&\quad\times A^{(i,j)}_R \Big( (u_L d_L) (\nu_{Lj} s_L)+  (u_L s_L) (\nu_{Lj} d_L)\Big) \nonumber \\
&&+\sum_{i,j=2,3} 2F(M_2, m^2_{{\tilde u}_{iL}},m^2_{{\tilde e}_{jL}})A^{(1,j)}_R \,{\overline m}_{u}{\overline m}_{d_j}  V_{u_i s} V^*_{u_i d_j} e^{i\varphi_i}  (u_L d_L) (\nu_{Lj} s_L) \nonumber \\
&& -\frac{{\overline m}^2_t {\overline m}_\tau V^*_{tb} e^{i\varphi_1}}{m^2_W \sin2\beta}\, F(\mu_H, m^2_{{\tilde t}_R},m^2_{{\tilde \tau}_R})\,\overline {A_R} \Big({\overline m}_d V_{ud} V_{ts} (u_R d_R)(\nu_\tau s_L) \nonumber \\
&&\quad\quad + {\overline m}_s V_{us} V_{td} (u_R s_R)(\nu_\tau d_L)\Big) \bigg] \label{dim6}
\eea
where $F(x,y,z)$ is the loop function, given by
\bea
F(x,y^2,z^2) = \frac{x}{(x^2-y^2)(y^2-z^2)(z^2-x^2)}\,\bigg[ x^2 y^2 \ln\Big(\frac{x^2}{y^2}\Big)+y^2 z^2\ln\Big(\frac{y^2}{z^2}\Big)+z^2x^2\ln\Big(\frac{z^2}{x^2}\Big)\bigg], \label{protonloop}
\eea
$A^{(i,j)}_R, \overline {A_R} $ are the renormalization factors, and ${\overline m}_{u_i}, {\overline m}_{d_i}$ are the running quark masses defined in the $\overline{{\rm DR}}$ scheme at the scale of $\mu=2\,{\rm GeV}$ \cite{SU5}.
Here, the flavor-diagonal squark and slepton masses correspond to $m^2_{{\tilde s}_L}=m^2_{{\tilde b}_L}$, $m^2_{{\tilde c}_L}=m^2_{{\tilde t}_L}$, and $m^2_{{\tilde\tau}_R}=m^2_{{\tilde\mu}_R}$, for the first two generation sparticles, and the electroweak symmetry leads to $m^2_{{\tilde s}_L}=m^2_{{\tilde c}_L}$ and $m^2_{{\tilde b}_L}=m^2_{{\tilde t}_L}$. However, we don't have to set $m^2_{{\tilde t}_R}=m^2_{{\tilde\tau}_R}$ for satisfying the bounds on FCNCs.

\begin{figure}[!t]
\centering
\includegraphics[width=0.45\textwidth,clip]{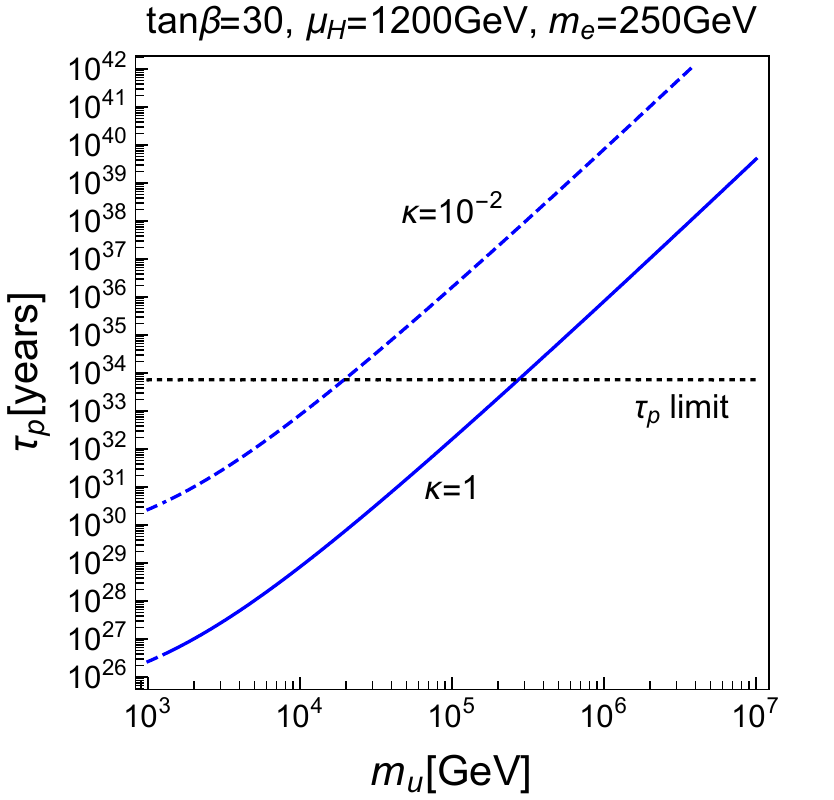}\,\,\,\,\,\,
\includegraphics[width=0.45\textwidth,clip]{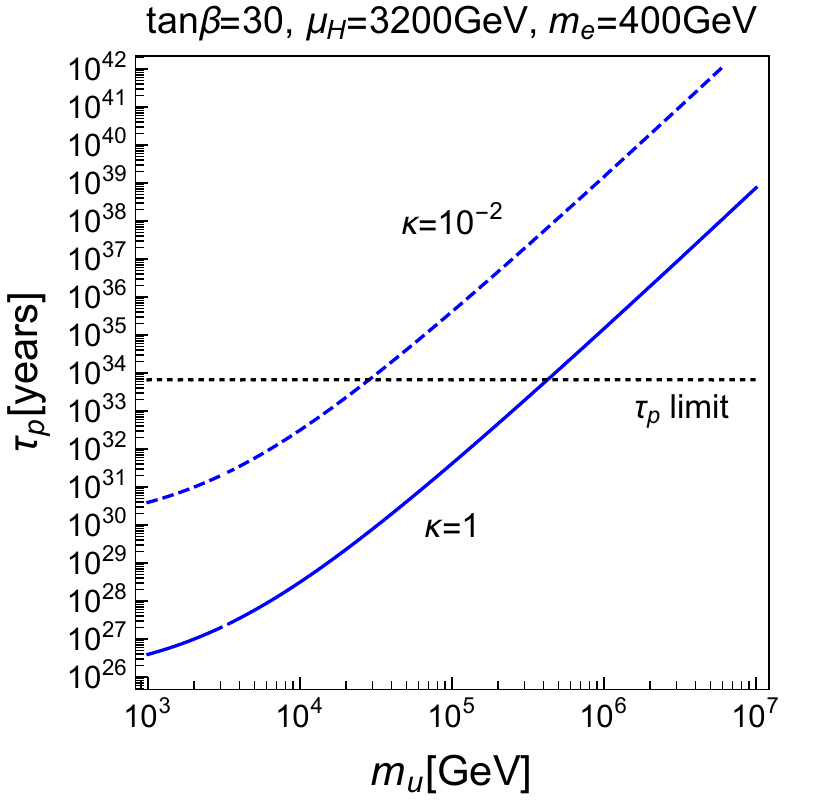} \\
\caption{(Left) Proton lifetime in years as a function of the squark mass, $m_{{\tilde u}_R}$ in units of GeV. We took $\kappa=1, 10^{-2}$ in blue solid and dashed lines, respectively. We chose $\tan\beta=30$ in common, $\mu_H=1200\,{\rm GeV}$, $m_{{\tilde e}_R}=250\,{\rm GeV}$ on left and  $\mu_H=3200\,{\rm GeV}$, $m_{{\tilde e}_R}=400\,{\rm GeV}$ on right.
}
\label{proton1}
\end{figure}

The slepton loops with charginos in the second line in eq.~(\ref{dim6}) are suppressed by the up-quark mass ${\overline m}_{u}$, so they are negligible even if $m^2_{{\tilde e}_{jL}}\ll m^2_{{\tilde u}_{iL}}$ is taken in the loop function. 
On the other hand, the stop and stau loops with Higgsinos in eq.~(\ref{dim6}) give rise to the dominant contribution to proton decay,  so we can find a correlation of the muon $g-2$ with the proton decay only if smuon and stau masses are related.
As discussed in the previous section, in the general gauge mediation, we can take the squark masses to be generation-independent and sufficiently large while keeping the slepton masses flavor universal and light.  Then, for smuon and stau of comparable masses, we can correlate between the muon $g-2$ with smuon loops and the proton decay with stau loops.

We remark that for the degenerate scalar superpartner masses in loops, the loop function in eq.~(\ref{protonloop}) becomes
\bea
F(x,y^2,y^2) = x\bigg[\frac{1}{y^2-x^2}-\frac{x^2}{(y^2-x^2)^2}\ln\Big(\frac{y^2}{x^2}\Big) \bigg],
\eea
recovering the results in Ref.~\cite{SU5,update}.

As a result,  the first two terms in eq.~(\ref{dim6}) are suppressed by either light quark masses or small CKM mixing angles, as compared to the third term in eq.~(\ref{dim6}).  Then, for $|\mu_H|\gtrsim M_2$ and $m_{{\tilde u}_{iL}}, m_{{\tilde d}_{iL}},m_{{\tilde t}_{R}}\gg m_{{\tilde e}_{iL}},m_{{\tilde e}_{iR}}, M_2, |\mu_H|$,  the loop function in the first term in eq.~(\ref{dim6}) is parametrically smaller than the one in the third term in eq.~(\ref{dim6}), so we can estimate the proton lifetime from Higgsino loops as
\bea
\tau(p\to K^+{\bar \nu}) \simeq 4\times 10^{35} \,{\rm years}\times \sin^4 2\beta \,\bigg(\frac{0.1}{\overline {A_R}}\bigg)^2 \bigg(\frac{[2F(\mu_H, m^2_{{\tilde t}_R},m^2_{{\tilde \tau}_R})]^{-1}}{10^2\,{\rm TeV}}\bigg)^2\,\bigg(\frac{M_{H_C}/\kappa}{10^{16}\,{\rm GeV}}\bigg)^2.
\eea
In comparison, we note that the current experimental limits are given by $\tau(p\to K^+{\bar \nu})>6.6\times 10^{33}\,{\rm yrs}$ \cite{SK}.

In Fig.~\ref{proton1}, we obtain the proton lifetime as a function of the squark mass, $m_{{\tilde u}_R}$ for varying the $\mu$-term and stau mass to $\mu_H=1200\,{\rm GeV}$, $m_{{\tilde e}_R}=250\,{\rm GeV}$ on left and  $\mu_H=3200\,{\rm GeV}$, $m_{{\tilde e}_R}=400\,{\rm GeV}$ on right. We also varied the suppression factor to $\kappa=1, 10^{-2}$ in blue solid and dashed lines, respectively. The black dotted line corresponds to the current limit on the proton decay mode, $p\to K^+{\bar \nu}$.
Here, since we assume that squark and slepton masses are generation-independent up to electroweak corrections, we don't distinguish flavors for squark and slepton masses.

\begin{figure}[!t]
\centering
\includegraphics[width=0.45\textwidth,clip]{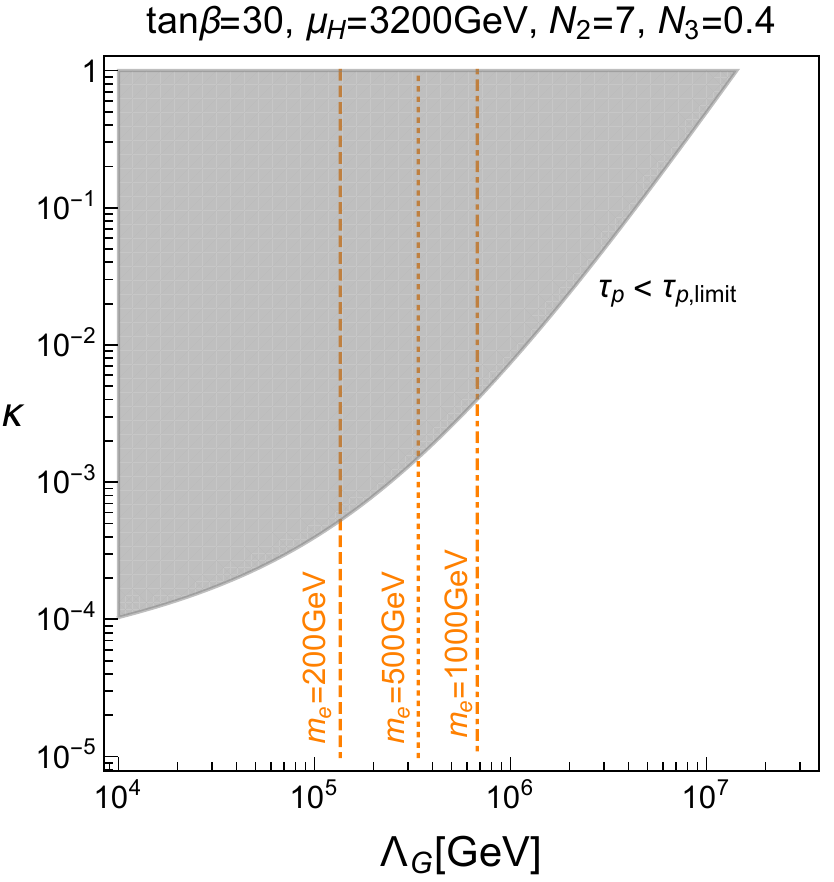}\,\,\,\,\,\,
\includegraphics[width=0.45\textwidth,clip]{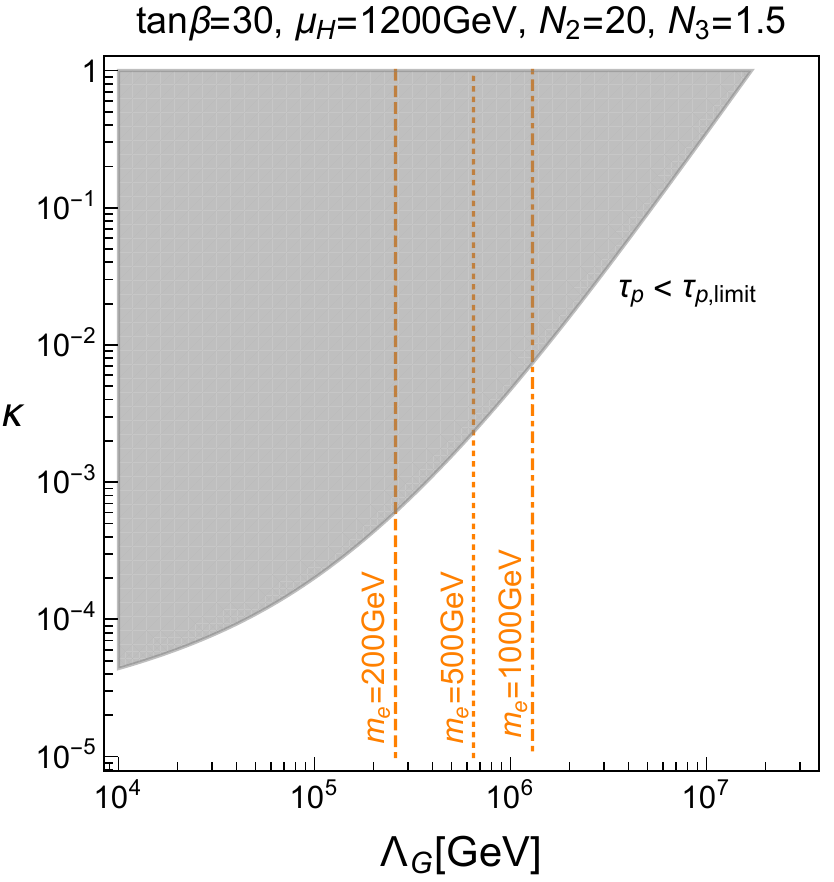} \\
\caption{Parameter space for the scale of general gauge mediation, $\Lambda_G$, and the suppression factor for the colored Higgsino Yukawa couplings, $\kappa$. The gray region is excluded by the bound on the proton lifetime for $p\to K^+ {\bar \nu}$. We showed the contours of the slepton mass, $m_{{\tilde e}_R}=200, 500, 1000\,{\rm GeV}$ in orange dashed, dotted and dot-dashed lines, respectively. We took $\tan\beta=30$ in common,  and $\mu_H=1200\,{\rm GeV}$, $N_2=7, N_3=0.4$ on left and $\mu_H=3200\,{\rm GeV}$, $N_2=20, N_3=1.5$ on right.
}
\label{proton2}
\end{figure}

We now make a direct connection of the proton lifetime to the parameters in general gauge mediation.
In Fig.~\ref{proton2}, we depict the parameter space for the scale of general gauge mediation, $\Lambda_G$, and the suppression factor, $\kappa$, which is excluded by the current bound on the proton lifetime. We took $\tan\beta=30$ for both plots and $\mu_H=1200\,{\rm GeV}$, $N_2=7, N_3=0.4$  (Model I in Table 1) on left and $\mu_H=3200\,{\rm GeV}$, $N_2=20, N_3=1.5$ (Model II in Table 1) on right. Orange dashed, dotted and dot-dashed lines correspond to the stau masses, $m_{{\tilde e}_R}=200, 500, 1000\,{\rm GeV}$, respectively. In both cases,  for squark masses of order $3-7\,{\rm TeV}$ in the benchmark models, we need a suppression factor, $\kappa\sim 10^{-4}-10^{-3}$, to satisfy the bound on the proton lifetime  and the slepton masses of order $100\,{\rm GeV}$ required for the muon $g-2$ anomaly. Therefore, the low-energy spectrum with split sparticles obtained in general gauge mediation is consistent with the muon $g-2$ anomaly and the strong bounds from the LHC, but it hints at a new mechanism beyond the minimal SU(5) GUTs for suppressing the Yukawa couplings of colored Higgsinos sufficiently for proton stability.

\begin{figure}[!t]
\centering
\includegraphics[width=0.45\textwidth,clip]{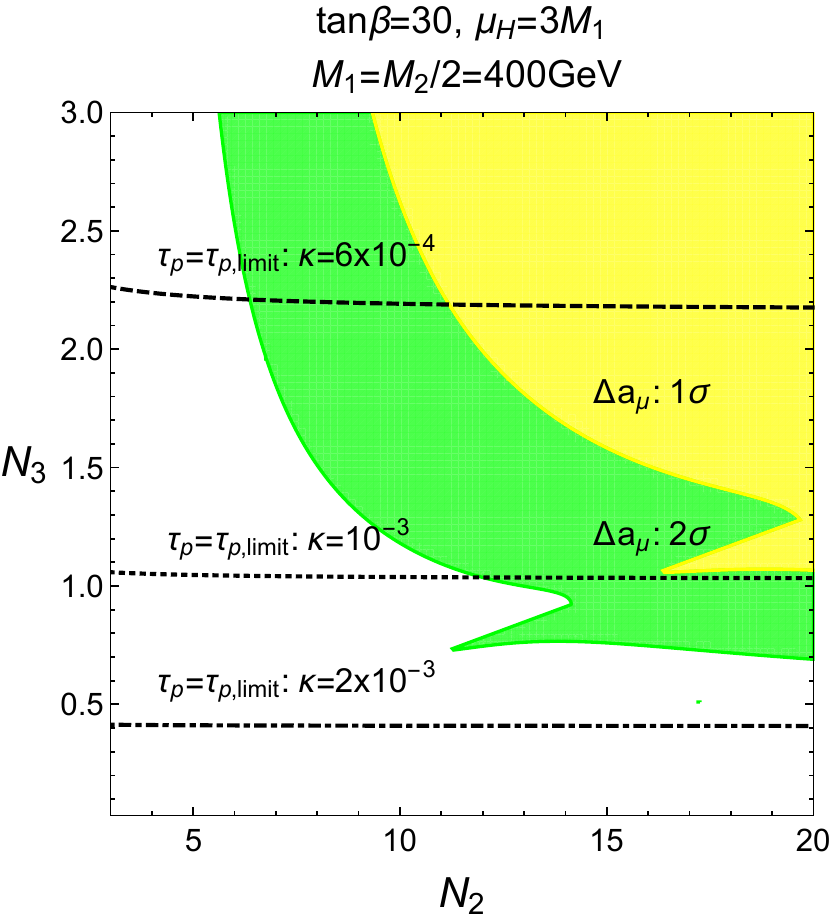}\,\,\,\,\,\,
\includegraphics[width=0.45\textwidth,clip]{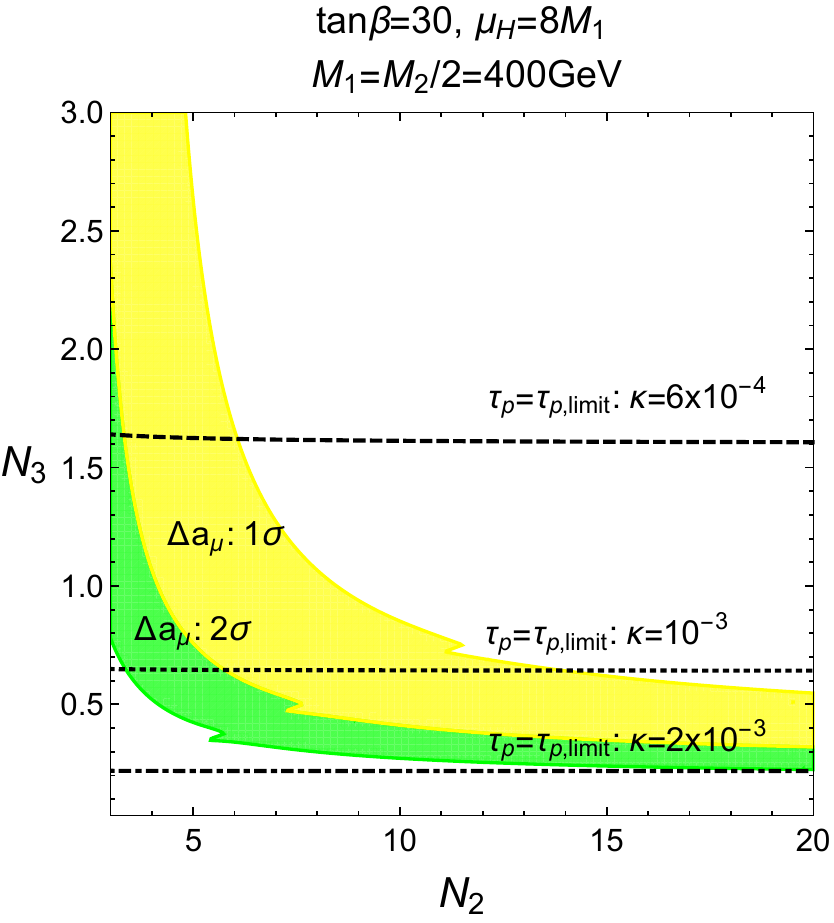} \\
\caption{Parameter space for $N_2$ and $N_3$, explaining the muon $g-2$ and satisfying the bound on the proton lifetime for $p\to K^+ {\bar \nu}$. We showed the parameter space saturating the bound on the proton lifetime for the suppression factor, $\kappa=6\times 10^{-4}, 10^{-3}, 2\times 10^{-3}$, in dashed, dotted and dot-dashed lines, respectively. We took $\tan\beta=30$, $M_1=M_2/2=400\,{\rm GeV}$ for both plots, and $\mu_H=3M_1$ on left and $\mu_H=8M_1$ on right. The color codes for yellow and green regions are the same as in Figs.~\ref{fig:g-2a}  or \ref{fig:g-2b}.
}
\label{proton3}
\end{figure}

In Fig.~\ref{proton3}, in order to make the correlation between the muon $g-2$ and the proton lifetime more explicit, in the parameter space for $N_2$ and $N_3$, namely, the effective number of doublet and triplet messenger fields in general gauge mediation, we show the region where the muon $g-2$ is explained as in Figs.~\ref{fig:g-2a}  or \ref{fig:g-2b} and overlay the contours for saturating with the bound on the proton lifetime for  $p\to K^+ {\bar \nu}$, for the suppression factor, $\kappa=6\times 10^{-4}, 10^{-3}, 2\times 10^{-3}$, in dashed, dotted and dot-dashed lines, respectively. The regions below the dashed, dotted and dot-dashed lines are consistent with the bound on the proton lifetime. On the other hand, the muon $g-2$ can be explained in yellow and green regions at the $1\sigma$ and $2\sigma$ levels, respectively. We took $\tan\beta=30$, $M_1=M_2/2=400\,{\rm GeV}$ for both plots, and $\mu_H=3M_1$ on left and $\mu_H=8M_1$ on right. We find that as far as $\kappa\lesssim 10^{-3}$, the muon $g-2$ can be readily explained within $1\sigma$, being consistent with the bound on the proton lifetime. For a larger $\mu_H$ as in the right plot of Fig.~\ref{proton3}, we can accommodate larger slepton masses to explain the muon $g-2$, namely, a smaller $N_3$, so the bound on the proton lifetime can be satisfied for a larger $\kappa$.

\section{Conclusions}

We considered the interplay of the muon $g-2$ anomaly and the proton decay in the SUSY SU(5) GUTs with generation-independent scalar soft masses at the messenger scale. In particular, in the scenarios of general gauge mediation with $\bf 5+{\bar 5}$ messenger fields, the messenger sector with doublet-triplet splitting transmits SUSY breaking to the visible sector such that squark and slepton masses are generation-independent but split already at the messenger scale. Thus, taking into account the perturbative unification of gauge couplings as well as the bounds from electroweak precision and vacuum stability bounds, we showed the parameter space in general gauge mediation to explain the muon $g-2$ anomaly with smuon and sneutrino loops while evading the strong bounds on squarks and gluinos from the LHC. 

From the relation between the gluino and bino masses, $M_3\simeq 6M_1$, which is the same as in the constrained MSSM model, we chose the bino mass $M_1$ to about $400\,{\rm GeV}$ to satisfy the LHC bounds on the gluino masses. In this case, we found that the lighter smuon is lighter than the lightest neutralino in the parameter space explaining the muon $g-2$ anomaly within $1\sigma$, but the lightest neutralino can be also the LSP at the $2\sigma$ level for the muon $g-2$. It is typical that the gravitino can be lighter than the LSP in MSSM in gauge mediation due to the messenger scale smaller than the Planck scale. Thus,  the gravitino can be a dark matter candidate whereas the LSP in MSSM is long-lived, decaying into the gravitino and an SM particle. Depending on whether the LSP in MSSM decays within the detector or not, the long-lived particle searches of displaced vertices or the standard SUSY searches with missing transverse momentum are applicable.

In benchmark models with hierarchical effective numbers of colored and non-colored messenger fields, we showed that the muon $g-2$ anomaly can be explained while the direct bounds for superparticles at the LHC are satisfied. We obtained the dominant Higgsino contributions to the proton decay, $p\to K^+{\bar\nu}$, with general generation-independent sparticle masses for squarks and sleptons appearing in loops. However, we showed that the scalar soft masses obtained in our model are not split enough, so we need an extra suppression factor for the Yukawa couplings of the colored Higgsinos for the proton stability, such as the Dirichlet boundary conditions on the colored Higgsinos in the extra dimension on orbifolds. We showed that the dimensionless suppression factor of order $10^{-4}-10^{-3}$ is sufficient to reconcile the slepton masses of a few $100\,{\rm GeV}$ needed for the muon $g-2$ anomaly with the unification.

\section*{Acknowledgments}

We are supported in part by Basic Science Research Program through the National
Research Foundation of Korea (NRF) funded by the Ministry of Education, Science and
Technology (NRF-2022R1A2C2003567 and NRF-2021R1A4A2001897). 
Sung-Bo Sim is supported by the Chung-Ang University Research Scholarship Grants in 2023.

\def\theequation{A.\arabic{equation}}

\setcounter{equation}{0}

\vskip0.8cm
\noindent
{\Large \bf Appendix A: Neutralino and chargino mixings}

Keeping the corrections from the electroweak symmetry breaking, we get the components of the neutralino mixing matrix in eq.~(\ref{neutralinomix}) approximately  as
\bea
N_{11}&=&1, \quad N_{12}=0, \\
N_{13}&=& \frac{(M_1\cos\beta+\mu_H \sin\beta)m_Z s_W}{\mu^2_H-M^2_1}, \\
N_{14}&=& - \frac{(M_1\sin\beta+\mu_H \cos\beta)m_Z s_W}{\mu^2_H-M^2_1},
\eea
\bea
N_{21}&=&0, \quad N_{22}=1, \\
N_{23}&=&- \frac{(M_2\cos\beta+\mu_H \sin\beta)m_Z c_W}{\mu^2_H-M^2_2}, \\
 N_{24}&=& \frac{(M_2\sin\beta+\mu_H \cos\beta)m_Z c_W}{\mu^2_H-M^2_2},
\eea
\bea
N_{31} &=& -\frac{1}{\sqrt{2}}\,\frac{m_Zs_W}{\mu_H-M_1}\,(\cos\beta+\sin\beta), \\
N_{32} &=& \frac{1}{\sqrt{2}}\,\frac{m_Zc_W}{\mu_H-M_2}\,(\cos\beta+\sin\beta), \\
N_{33}&=&-N_{34}=\frac{1}{\sqrt{2}},
\eea
and
\bea
N_{41} &=& \frac{1}{\sqrt{2}}\,\frac{m_Z s_W}{\mu_H+M_1}\,(\cos\beta-\sin\beta), \\
N_{42} &=& -\frac{1}{\sqrt{2}}\,\frac{m_Z c_W}{\mu_H+M_2}\,(\cos\beta-\sin\beta), \\
N_{43}&=&N_{44}=\frac{1}{\sqrt{2}}.
\eea

Similarly, we also obtain the components of the chargino mixing matrices in eq.~(\ref{charginomix}) approximately as
\bea
U_{L,11}&=&U_{L,22}=1, \\
U_{L,12}&=&-U_{L,21}=-\frac{\sqrt{2}m_W(M_2\cos\beta+\mu_H \sin\beta)}{\mu^2_H-M^2_2},
\eea
and
\bea
U_{R,11}&=&U_{R,22}=1, \\
U_{R,12}&=&-U_{R,21}=-\frac{\sqrt{2}m_W(M_2\sin\beta+\mu_H \cos\beta)}{\mu^2_H-M^2_2}.
\eea

\def\theequation{B.\arabic{equation}}

\setcounter{equation}{0}

\vskip0.8cm
\noindent
{\Large \bf Appendix B: Effective interactions for sleptons}

Keeping the corrections from the electroweak symmetry breaking to the neutralino and chargino mass matrices, we get the approximate effective couplings for the sleptons in eq.~(\ref{effsleptons}) explicitly as
\bea
B^{L}_{1,1}&\simeq& \cos\theta_{\tilde\mu}+ \frac{1}{\sqrt{2}g'} f_\mu\sin\theta_{\tilde\mu} \cdot\frac{(M_1c_\beta+\mu_Hs_\beta)m_Zs_W}{\mu^2_H-M^2_1}, \\ 
B^{L}_{1,2}&\simeq& \sin\theta_{\tilde\mu}+ \frac{1}{\sqrt{2}g'} f_\mu\cos\theta_{\tilde\mu} \cdot \frac{(M_1c_\beta+\mu_Hs_\beta)m_Zs_W}{\mu^2_H-M^2_1}, \\
B^{L}_{2,1}&\simeq& - \frac{1}{\sqrt{2}g'} f_\mu\sin\theta_{\tilde\mu}\cdot \frac{(M_1c_\beta+\mu_Hs_\beta)m_Zc_W}{\mu^2_H-M^2_2},  \\
B^{L}_{2,2}&\simeq& - \frac{1}{\sqrt{2}g'} f_\mu\cos\theta_{\tilde\mu}\cdot \frac{(M_1c_\beta+\mu_Hs_\beta)m_Zc_W}{\mu^2_H-M^2_2}, \\
B^{L}_{3,1}&\simeq&\frac{1}{2g'} f_\mu\sin\theta_{\tilde\mu} -\frac{1}{\sqrt{2}} \frac{m_Z s_W}{\mu_H-M_1}(c_\beta+s_\beta)\cos\theta_{\tilde\mu}, \\
B^{L}_{3,2} &\simeq&\frac{1}{2g'} f_\mu\cos\theta_{\tilde\mu} -\frac{1}{\sqrt{2}} \frac{m_Z s_W}{\mu_H-M_1}(c_\beta+s_\beta)\sin\theta_{\tilde\mu}, \\
B^{L}_{4,1}&\simeq&\frac{1}{2g'}  f_\mu\sin\theta_{\tilde\mu}+\frac{1}{\sqrt{2}}\frac{m_Zs_W}{\mu_H+M_1}(c_\beta-s_\beta)\cos\theta_{\tilde\mu} , \\
B^{L}_{4,2} &\simeq& \frac{1}{2g'}  f_\mu\cos\theta_{\tilde\mu}+\frac{1}{\sqrt{2}}\frac{m_Zs_W}{\mu_H+M_1}(c_\beta-s_\beta)\sin\theta_{\tilde\mu},
\eea
and
\bea
B^{R}_{1,1}&\simeq& \frac{1}{2}\sin\theta_{\tilde\mu}- \frac{1}{\sqrt{2}g'} f_\mu\cos\theta_{\tilde\mu} \cdot \frac{(M_1c_\beta+\mu_H s_\beta)m_Z s_W}{\mu^2_H-M^2_1}, \\
B^{R}_{1,2}&\simeq& -\frac{1}{2}\cos\theta_{\tilde\mu}-\frac{1}{\sqrt{2}g'} f_\mu\sin\theta_{\tilde\mu}\cdot  \frac{(M_1c_\beta+\mu_H s_\beta)m_Z s_W}{\mu^2_H-M^2_1}, \\
B^{R}_{2,1}&\simeq& \frac{1}{2}\sin\theta_{\tilde\mu}\cot\theta_W+\frac{1}{\sqrt{2}g'} f_\mu\cos\theta_{\tilde\mu} \cdot \frac{(M_2c_\beta+\mu_H s_\beta)m_Z c_W}{\mu^2_H-M^2_2}, \\
B^{R}_{2,2}&\simeq& -\frac{1}{2}\cos\theta_{\tilde\mu}\cot\theta_W+\frac{1}{\sqrt{2}g'} f_\mu\sin\theta_{\tilde\mu}\cdot  \frac{(M_2c_\beta+\mu_H s_\beta)m_Z c_W}{\mu^2_H-M^2_2}, \\
B^{R}_{3,1}&\simeq&-\frac{1}{2g'}\,f_\mu\cos\theta_{\tilde\mu}-\frac{1}{2\sqrt{2}}\sin\theta_{\tilde\mu}(c_\beta+s_\beta)\bigg(\frac{m_Zs_W}{\mu_H-M_1}-\frac{m_Zc^2_W/s_W}{\mu_H-M_2}\bigg), \\
B^{R}_{3,2} &\simeq&-\frac{1}{2g'}\,f_\mu\sin\theta_{\tilde\mu}+\frac{1}{2\sqrt{2}}\cos\theta_{\tilde\mu}(c_\beta+s_\beta)\bigg(\frac{m_Zs_W}{\mu_H-M_1}-\frac{m_Zc^2_W/s_W}{\mu_H-M_2}\bigg), \\
B^{R}_{4,1}&\simeq&-\frac{1}{2g'} f_\mu\cos\theta_{\tilde\mu}+ \frac{1}{2\sqrt{2}}\sin\theta_{\tilde\mu} (c_\beta-s_\beta)\bigg(\frac{m_Zs_W}{\mu_H+M_1}-\frac{m_Zc^2_W/s_W}{\mu_H+M_2}\bigg), \\
B^{R}_{4,2} &\simeq&-\frac{1}{2g'} f_\mu\sin\theta_{\tilde\mu} -\frac{1}{2\sqrt{2}}\cos\theta_{\tilde\mu} (c_\beta-s_\beta)\bigg(\frac{m_Zs_W}{\mu_H+M_1}-\frac{m_Zc^2_W/s_W}{\mu_H+M_2}\bigg).
\eea

\end{document}